\documentclass[onefignum,onetabnum]{siamart190516}
\providecommand{\tabularnewline}{\\}

\usepackage[OT1]{fontenc}
\usepackage[latin9]{inputenc}
\usepackage{array}
\usepackage{mathtools}
\usepackage{multirow}
\usepackage{breakurl}
\usepackage{physics, siunitx, amsmath, amssymb, mathtools,  pgfplots, caption, subcaption, booktabs}
\usepackage[version=4]{mhchem}
\usepackage{comment}
\usepackage{placeins}
\pgfplotsset{compat=1.13}
\usepackage{todonotes}
\usepackage{ragged2e}
\newcommand{\beq}{\begin{equation}}
\newcommand{\eeq}{\end{equation}}
\newcommand{\cspm}{c^{\textrm{SPM}}}
\newcommand{\cc}{c}
\newcommand{\GG}{{\mathcal G}}
\newcommand{\pphi}{\Phi}
\newcommand{\ft}{}
\renewenvironment{cases}{%
 \begin{dcases}%
}{%
 \end{dcases}%\kern-\nulldelimiterspace%
}

\def\XXint#1#2#3{{\setbox0=\hbox{$#1{#2#3}{\int}$ }
\vcenter{\hbox{$#2#3$ }}\kern-.6\wd0}}

\makeatother

\usepackage{amsfonts}
\usepackage{graphicx}
\usepackage{epstopdf}
\usepackage{algorithmic}
\ifpdf
  \DeclareGraphicsExtensions{.eps,.pdf,.png,.jpg}
\else
  \DeclareGraphicsExtensions{.eps}
\fi

\headers{Electrode heterogeneity in lithium-ion batteries}{T.L. Kirk, J. Evans,  C.P. Please and  S.J. Chapman}

\title{Modelling electrode heterogeneity in lithium-ion batteries: unimodal
and bimodal particle-size distributions\thanks{Submitted to the editors June 4, 2020.
\funding{TLK, CPP and SJC were supported by the Faraday Institution Multi-Scale Modelling (MSM) project, grant number EP/S003053/1.}}}

\author{Toby L. Kirk\thanks{Mathematical Institute, University of Oxford, Andrew Wiles Building, Woodstock Road, Oxford OX2 6GG, UK (\email{toby.kirk@maths.ox.ac.uk}, \email{please@maths.ox.ac.uk}, \email{chapman@maths.ox.ac.uk}).}
  \and Jack Evans\thanks{Department of Materials, University of Oxford, 21 Banbury Road, Oxford OX2 6HT, UK (\email{jack.evans@seh.ox.ac.uk}).}
  \and Colin P.~Please\footnotemark[2]
\and S.~Jonathan Chapman\footnotemark[2]}

\usepackage{amsopn}

\ifpdf
\hypersetup{
  pdftitle={Modelling electrode heterogeneity in lithium-ion batteries: unimodal
and bimodal particle-size distributions},
  pdfauthor={Toby L. Kirk, Jack Evans,  C.P. Please and  S.J. Chapman}
}
\fi

\begin{document}

\maketitle

\begin{abstract}
In mathematical models of lithium-ion batteries, the highly heterogeneous
porous electrodes are frequently approximated as comprising spherical
particles of uniform size, leading to the commonly-used single-particle 
model (SPM) when transport in the electrolyte is assumed to be fast.
Here  electrode heterogeneity is modelled by extending this to 
a distribution of particle sizes.
Unimodal and bimodal particle-size distributions
(PSD) are considered. For a
unimodal PSD, the effect of the spread of the distribution on the
cell dynamics is investigated, and choice of effective particle radius when approximating by an SPM assessed. Asymptotic techniques are used to derive a correction to the
SPM valid for narrow, but realistic, PSDs. In addition, it is shown that
the heterogeneous internal states of all particles (relevant when modelling degradation, for example) can be efficiently
computed after-the-fact.
For a bimodal PSD, the results are well approximated by a double-particle
model (DPM), with one size representing each mode. Results
for lithium iron phosphate with a bimodal PSD show that the DPM
captures an experimentally-observed double-plateau in the discharge curve, suggesting it is
entirely due to bimodality.
\end{abstract}

% REQUIRED

% REQUIRED

\section{Introduction}

Lithium-ion batteries are rechargeable energy storage devices that
are ubiquitous in consumer electronics due to their high energy density,
long lifespan, and a low self-discharge rate compared to other batteries
\cite{Blomgren2017}. In recent years, they are increasingly being
employed in off-grid storage and electric vehicles, with demand predicted
to increase from 45 GWh per year in 2015 to 390 GWh per year by 2030
\cite{Zubi2018}. This necessitates urgent improvements in lithium-ion
battery performance, in particular, their safety, lifespan and capacity.

A lithium-ion battery typically consists of many single electrochemical
cells, with the principal components of each cell being a positive
electrode, negative electrode, and a liquid electrolyte. The electrodes
are porous materials comprising a collection of microscopic electrode
particles adhered together using a polymer binder, with the pores filled with electrolyte. Lithium atoms are
stored within the electrode particles and, during a discharge, a reaction
at their surface occurs whereby lithium deintercalates (is extracted)
from the negative electrode forming a lithium-ion, Li$^{+}$, and
a free electron. The ion travels through the electrolyte and intercalates
(is inserted) into the positive electrode, and the free electron travels
via an external circuit producing the electrical current. During a
charge, this process happens in reverse. 

 The multiscale nature
of battery processes has lead to modelling and simulation on scales from individual atoms up to a full battery\textemdash for a recent review
of the various scales and complexities see \cite{Franco2013}.
%\cite{Ramadesigan2012}
The present work focusses on the scale from an electrode particle to a single cell. The pioneering modelling at this scale, referred to
as porous electrode theory, was developed by the group of Newman \cite{DoyleFullerNewman1993,Fuller1994,NewmanBook}
where a macroscale (i.e. cell scale) model for potentials and lithium
concentrations is coupled at each location to a microscale (i.e.
particle scale) problem to accurately capture the surface reactions.
This approach, in particular the Doyle\textendash Fuller\textendash Newman
(DFN) model \cite{DoyleFullerNewman1993}, has since been justified
using asymptotic homogenisation \cite{Richardson2012}. 

When used for parameter estimation \cite{Bizeray2019}, or to optimise
cell design \cite{Cheng2019}, or when more macroscale dimensions
are present \cite{Yi2013}, DFN-type models can be prohibitively
expensive to solve numerically. Therefore, many simplifications have
been considered (see \cite{Jokar2016} for a recent review), the most
common being the Single-Particle Model (SPM) \cite{Ning2004,PerezMoura2016,Moura2017}
where it is assumed that all active particles in an electrode are
spheres of the same size and behave identically, and only a single
representative particle is modelled for each electrode. It has been
shown that the SPM can be obtained systematically from the DFN in
various asymptotic limits, e.g. fast transport of ions in the  electrolyte \cite{Marquis2019} or an open circuit potential that is sufficiently
non-flat \cite{Richardson2019}.
\begin{comment}
An advantage of asymptotic
expansions is that higher order corrections can be derived systematically
to improve accuracy with only a small additional computational effort\textemdash see
the SPM with electrolyte \cite{Marquis2019}. Another recent study
to employ asymptotic simplification is Moyles \emph{et al.} \cite{Moyles2019},
who reduced a porous electrode model to a pair of ODEs while maintaining
excellent agreement with experiment.
\end{comment}

An important feature of lithium-ion battery electrodes that is typically
neglected is the heterogeneity of the electrode microstructure. Almost
all modelling studies, except those discussed below, assume  for computational simplicity that electrode
particles are all the same shape and size.
In reality, particles of many different sizes and shapes are present
and, for a given shape, can be quantified by the particle-size distribution
(PSD). The PSD of an electrode can be readily determined experimentally
using a variety of techniques \cite{AllenBook,MalvernWP}, and control
of its shape has been demonstrated in manufacturing \cite{Fey2009,Drezen2007}.
It is well known that particle size has a significant effect on capacity
and degradation rates, with smaller particles providing better performance
\cite{Fey2009}. However, different particle sizes experience
different current densities which may result
in different degradation rates \cite{Goers2011}. In addition, the
PSD itself changes as the battery is cycled \cite{Zavalisetlal13},
with one explanation being particle cracking and agglomeration. Therefore,
the PSD of an electrode not only affects performance, but capturing
it in mathematical models will be essential in the accurate modelling
of degradation.

There have been several studies that have included multiple particle
sizes in battery models \cite{Darling1997,Srinivasan2004,Farkhondeh2011,Farkhondeh2014,Majdabadi2015,Taleghani2017}.
For the electrode material lithium iron phosphate (LiFePO$_{4}$),
additional particle sizes have been considered in an attempt to better
quantitatively fit discharge curves from experiment \cite{Srinivasan2004,Farkhondeh2011,Farkhondeh2014},
and possibly explain hysteresis and memory effects \cite{Kondo2018}.
The most relevant study is Farkhondeh \emph{et al.} \cite{Farkhondeh2011},
who presented an extension of the SPM to a Many-Particle Model (MPM)
to account for a PSD they determined from electron microscopy. However,
the PSD was approximated by only three particle sizes, with the smallest
size set to the median size and the larger two adjusted to fit their
MPM to experimental results. The agreement was limited to low (dis)charge
rates, and the model was later extended to a DFN-type model with three
particle sizes at every macroscale location \cite{Farkhondeh2014}.

The effect of the shape and spread of the PSD on battery behaviour
has received relatively little attention.
Its impact  on electrochemical impedance
has been considered by \cite{Meyers2000,Song2013}, but only R\"oder
\emph{et al.} \cite{Roder2016.} have investigated its impact on battery performance. Using an MPM similar
to that of \cite{Farkhondeh2011}, they showed that increasing the
spread of a unimodal (Weibull) PSD decreases the electrode capacity
available for discharge of graphite electrodes. This can be predicted
well by an SPM with particle size equal to the surface-area or volume
moment means of the PSD, the choice of mean depending on the parameter
regime. However, they use an analytical form of the open circuit potential
(OCP) which depends weakly (logarithmically) on the state of charge.
Hence, the impact of the PSD throughout a discharge is not observable
or presented in \cite{Roder2016.}.

The present work has two main motivations: (i) present a detailed
investigation of the effect of the PSD, not only on the total capacity
but on the electrode dynamics throughout a (dis)charge, accounting
for the nonlinear effects of the OCP that have been previously neglected;
(ii) reduce the model complexity encountered when using the PSD to
more easily account for these effects in battery models. For aim (i),
we present numerical solutions of an MPM for a wide class of continuous
PSDs, including unimodal and bimodal cases.
%(a continuous bimodal PSD
%has never been considered before in a battery model, to our knowledge).
For aim (ii), we assess approximating the problem with a single- or
double-particle model (SPM or DPM) for various choices of effective particle
size, and derive a new effective particle
size from the PSD to best predict the capacity
and the electrode behaviour near the end of a (dis)charge. To account
for the spread of a unimodal PSD throughout a discharge, we use asymptotic
techniques in the limit of narrow distributions to derive corrections
to the SPMs, consisting of 3 additional ``correction particles''.
Results are presented mainly for a graphite half-cell, the most common
anode material and one with a highly nonlinear OCP. However, we also
present experimental results for LiFePO$_{4}$ cathodes with bimodal
PSDs, and show that our DPM captures several key features, suggesting
they are entirely due to the presence of two modes in the PSD.

The structure of the paper is as follows. In \S\ref{sec:Problem-Formulation} we describe the
electrochemical model and the nondimensionalisation. In \S\ref{sec:Unimodel-PSDs}, we focus
on unimodal PSDs, their reduction to SPMs and associated asymptotic corrections,
followed by comparisons to full numerical solutions. In \S\ref{sec:Bimodal-Particle-Size-Distributi}, we consider
bimodal PSDs and the reduction to a DPM, followed by a comparison
to experimental results for LiFePO$_{4}$. Finally, the conclusions
are given in \S\ref{sec:Conclusions}.

\section{Problem Formulation}
\label{sec:Problem-Formulation}

\subsection{The Many-Particle Model}

The model we will consider is the Many-Particle Model (MPM), which
is similar to that in \cite{Farkhondeh2011,Farkhondeh2014,Zavalisetlal13,Roder2016.},
and is an extension of the commonly used Single-Particle Model (SPM)
to include particles of more than one size. The SPM can be derived
as a limit of the Doyle-Fuller-Newman (DFN) porous electrode model
\cite{DoyleFullerNewman1993} when the applied current is small \cite{Marquis2019}.
The MPM used here is also the corresponding small-current limit of
a DFN-type model with many particle sizes at each macroscale location,
with a similar derivation.
\begin{figure}
\begin{centering}
\includegraphics[width=0.75\textwidth]{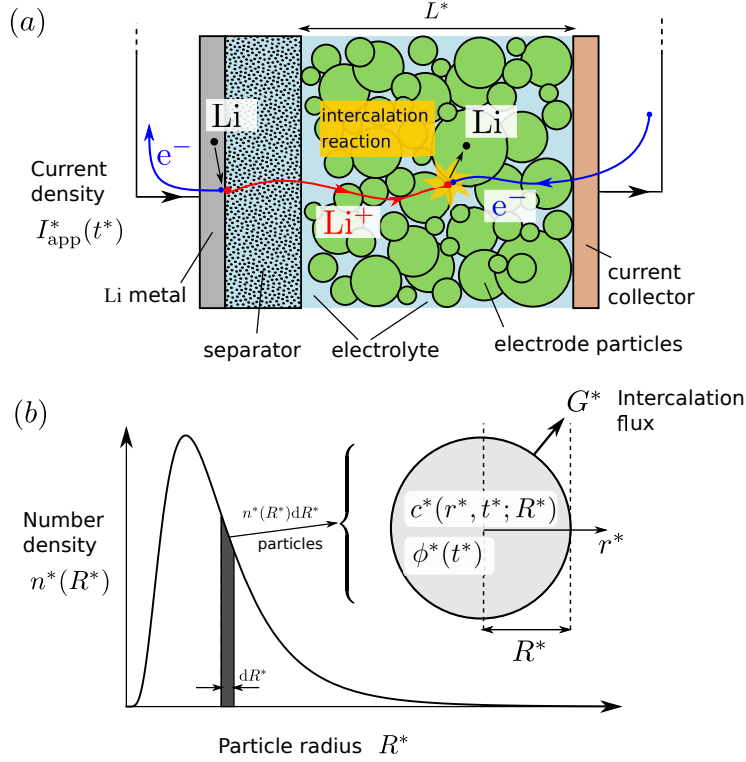}
\par\end{centering}
\caption{\label{fig:MPM-schematic}($a$) The half cell under
consideration; ($b$) the dimensional many-particle model (MPM)
for a typical unimodal (log-normal) number density $n^{*}(R^{*})$.}
\end{figure}

We will consider a half-cell geometry, consisting of a single porous
electrode immersed in a liquid electrolyte and separated from a lithium-metal
electrode by an insulating porous separator, as depicted in Fig.~\ref{fig:MPM-schematic}$(a)$.
In the small-current limit, the transport of lithium ions in the electrolyte
is fast so they remain at a constant uniform concentration $c_{e}^{*}$
there. Here we use stars to denote that a quantity is dimensional.
Only the lithium transport in the active material of the electrode,
and the electrochemical reaction on the electrode\textendash electrolyte
interfaces, need to be modelled. First we describe the particle geometry
and distribution, then state the electrochemical model.

\subsubsection{Particle-Size Distribution (PSD)}

The active particles that make up the porous electrode are assumed
to be spherical but of a range of different sizes. In general, we
assume there is a continuous distribution of particles of radii in
the range $0<R^{*}<\infty$, with the number of particles (per unit
electrode volume) with radii between $R^{*}$ and $R^{*}+\mathrm{d}R^{*}$
given by $n^{*}(R^{*})\mathrm{d}R^{*}$ \cite{RamkrishnaBook}. The physical location of
the particles in the electrodes is irrelevant because  ion transport
through the electrolyte is assumed instantaneous, and thus all particles
of a given radius $R^{*}$ behave identically\textemdash see Fig.
\ref{fig:MPM-schematic}$(b)$. The particles could be, for example,
distributed uniformly throughout the electrode or graded spatially
by size, and the model presented here is applicable in both cases.

The surface area $A^{*}(R^{*})$ and volume $V^{*}(R^{*})$ of a particle
of radius $R^{*}$ are, for spheres, given by
\begin{align}
A^{*}(R^{*}) & =4\pi R^{*2}, & V^{*}(R^{*}) & =\frac{4}{3}\pi R^{*3},\label{eq:A_and_V}
\end{align}
from which an area density $a^{*}(R^{*})$ and volume density $v^{*}(R^{*})$
can be defined,
\begin{align}
a^{*}(R^{*}) & =A^{*}(R^{*})n^{*}(R^{*}), & v^{*}(R^{*}) & =V^{*}(R^{*})n^{*}(R^{*}),\label{eq:area_density}
\end{align}
corresponding to the area and volume of all particles of radius $R^{*}$,
respectively. The total particle number, surface area, and volume
(each per unit  volume of electrode) are given by
\begin{align}
n_{\mathrm{total}}^{*} & =\int_{0}^{\infty}n^{*}(R^{*})\mathrm{d}R^{*}, & a_{\mathrm{total}}^{*} & =\int_{0}^{\infty}a^{*}(R^{*})\mathrm{d}R^{*}, & v_{\mathrm{total}} & =\int_{0}^{\infty}v^{*}(R^{*})\mathrm{d}R^{*}.\label{eq:density_totals}
\end{align}
The quantity $a_{\mathrm{total}}^{*}$ is known as the specific
surface area or Brunauer\textendash Emmett\textendash Teller (BET)
surface area, while $v_{\text{total}}$ is dimensionless and corresponds
to the volume fraction of active material, often denoted by $\epsilon_{s}$. We can normalise the densities
$n^{*}$, $a^{*}$, and $v^{*}$ by their totals to give fraction
densities
\begin{comment}
\begin{align}
f_{n}^{*}(R^{*}) & =\frac{n^{*}(R^{*})}{n_{\mathrm{total}}^{*}}=\frac{n^{*}(R^{*})}{\int_{0}^{\infty}n^{*}(R^{*})\mathrm{d}R^{*}},\\
f_{a}^{*}(R^{*}) & =\frac{a^{*}(R^{*})}{a_{\mathrm{total}}^{*}}=\frac{a^{*}(R^{*})}{\int_{0}^{\infty}a^{*}(R^{*})\mathrm{d}R^{*}}=\frac{R^{*2}f_{n}^{*}(R^{*})}{\int_{0}^{\infty}R^{*2}f_{n}^{*}(R^{*})\mathrm{d}R^{*}},\\
f_{v}^{*}(R^{*}) & =\frac{v^{*}(R^{*})}{v_{\mathrm{total}}}=\frac{v^{*}(R^{*})}{\int_{0}^{\infty}v^{*}(R^{*})\mathrm{d}R^{*}}=\frac{R^{*3}f_{n}^{*}(R^{*})}{\int_{0}^{\infty}R^{*3}f_{n}^{*}(R^{*})\mathrm{d}R^{*}},\label{eq:rho_v_star}
\end{align}
\end{comment}
\beq
f_{n}^{*}(R^{*})  =\frac{n^{*}(R^{*})}{n_{\mathrm{total}}^{*}}, \qquad
f_{a}^{*}(R^{*})  =\frac{a^{*}(R^{*})}{a_{\mathrm{total}}^{*}},
\qquad
f_{v}^{*}(R^{*})  =\frac{v^{*}(R^{*})}{v_{\mathrm{total}}},\label{eq:f_star}
\eeq
satisfying $\int_{0}^{\infty}f_{i}^{*}(R^{*})\mathrm{d}R^{*}=1$ for
each of $i=n,a,v$. The quantity $f_{n}^{*}(R^{*})$ can be interpreted
as the \emph{radius distribution}, i.e., the probability density function
of a particle having radius $R^{*}$ if randomly sampled from the
population. However, $f_{a}^{*}(R^{*})$ and $f_{v}^{*}(R^{*})$
are not the \emph{area} and \emph{volume distributions} in this sense, and should
instead be thought of as new radius distributions created by weighting
$f_{n}^{*}(R^{*})$ by the area and volume, respectively, and then
renormalising. Hence we shall refer to them as the area- and volume-weighted
radius distributions; they are commonly used in place of $f_{n}^{*}(R^{*})$,
depending on the experimental technique used to measure the PSD.

As the PSD is completely described by the number density $n^{*}(R^{*})$,
we may specify the total number of particles, $n_{\mathrm{total}}^{*}$,
and how they are distributed by radius, $f_{n}^{*}(R^{*})$, with
the remaining quantities in (\ref{eq:area_density})-(\ref{eq:f_star})
following from their definitions. One could, however, specify any
one of (\ref{eq:density_totals}) instead of $n_{\mathrm{total}}^{*}$,
and we will choose $v_{\mathrm{total}}$, which is easiest to determine
experimentally. It also influences the maximum theoretical capacity
of the electrode, which we will keep fixed  as we vary the shape and
spread of the distribution. Substituting $a^{*}(R^{*})=n_{\mathrm{total}}^{*}4\pi R^{*2}f_{n}^{*}(R^{*})$
and $v^{*}(R^{*})=n_{\mathrm{total}}^{*}\frac{4}{3}\pi R^{*3}f_{n}^{*}(R^{*})$
into (\ref{eq:density_totals}) and solving for $n_{\mathrm{total}}^{*}$
and $a_{\mathrm{total}}^{*}$ gives
\beq
n_{\mathrm{total}}^{*}  =\frac{v_{\mathrm{total}}}{\frac{4}{3}\pi\int_{0}^{\infty}R^{*3}f_{n}^{*}(R^{*})\mathrm{d}R^{*}},\qquad
a_{\mathrm{total}}^{*}  =\frac{3v_{\mathrm{total}}\int_{0}^{\infty}R^{*2}f_{n}^{*}(R^{*})\mathrm{d}R^{*}}{\int_{0}^{\infty}R^{*3}f_{n}^{*}(R^{*})\mathrm{d}R^{*}}.
\eeq
It will be useful to introduce the raw moments
\[
m_{i,j}^{*}=\int_{0}^{\infty}R^{*j}f_{i}^{*}(R^{*})\mathrm{d}R^{*},\qquad i=n,a,v,\quad j=1,2,\ldots
\]
as well as the means (first raw moment) and variances
\beq
\bar{R}_{i}^{*}  =\mathbb{E}_{i}[R^{*}] =m_{i,1}^{*}, \qquad
\sigma_{i}^{*2}  =\mathbb{E}_{i}[(R^{*}-\bar{R}_{i}^{*})^{2}]  =m_{i,2}^{*}-m_{i,1}^{*2}, \qquad i  =n,a,v.\label{eq:variance_i_star}
\eeq
The mean $\bar{R}_{n}^{*}$ corresponds to the average particle radius.
For further discussion about the means $\bar{R}_{i}^{*}$ and their
physical meaning see \S\ref{subsec:Single-particle-models}.

\subsubsection{Electrochemical model}

The transport of lithium within a particle of radius $R^{*}$
is assumed to be spherically symmetric, and modelled by Fickian
diffusion,
\begin{align}
\frac{\partial c^{*}}{\partial t^{*}} & =D^{*}\frac{1}{r^{*2}}\frac{\partial}{\partial r^{*}}\left(r^{*2}\frac{\partial c^{*}}{\partial r^{*}}\right),\qquad\text{for }0<r^{*}<R^{*},\label{eq:dimensional_diffusion_eq}\\
c^{*} & \equiv c_{\text{init}}^{*},\qquad\text{at }t^{*}=0,
\end{align}
where $c^{*}(r^{*},t^{*};R^{*})$ is the concentration of lithium
in the solid electrode material, with initial uniform concentration
$c_{\text{init}}^{*}$, and $D^{*}$ is the solid-state diffusion
coefficient. There is
%regularity at the centre of the particles,
%enforced
%by the condition
%\begin{equation}
%\frac{\partial c^{*}}{\partial r^{*}}=0,\qquad\text{at %}r^{*}=0,\label{eq:dimensional_centre_BC}
%\end{equation}
%and
a surface flux, i.e. lithium (de)intercalation into (out of)
the particles,
\begin{equation}
-D^{*}\frac{\partial c^{*}}{\partial r^{*}}=G^{*},\qquad\text{at }r^{*}=R^{*},\label{eq:dimensional_surface_BC}
\end{equation}
modelled by standard Butler\textendash Volmer kinetics,
\begin{equation}
G^{*}=\frac{k^{*}}{F^{*}}(c^{*})^{1/2}(c_{\text{max}}^{*}-c^{*})^{1/2}(c_{e}^{*})^{1/2}\sinh\left(\frac{F^{*}}{2R_{g}^{*}T^{*}}\eta^{*}\right),\qquad\text{at }r^{*}=R^{*}.\label{eq:dimensional_Butler_Volmer}
\end{equation}
Here $F^{*}$ is Faraday's constant, $R_{g}^{*}$ is the universal
gas constant, $T^{*}$ is the temperature (assumed constant), $k^{*}$
is a reaction rate coefficient, $c_{\text{max}}^{*}$ is the maximum
lithium concentration in the electrode material,
%$c_{s}^{*}(t;R)=c^{*}|_{r^{*}=R^{*}}$
%is the surface concentration,
$c_{e}^{*}$ is the concentration of lithium ions in the electrolyte (assumed constant), and the transfer coefficients for the
anodic and cathodic reactions are taken to be one half.
The function $\eta^{*}=\Delta\phi^{*}(t^{*})-U^{*}(c_{s}^{*},c_{e}^{*})$ is the surface overpotential, where $\Delta\phi^{*}(t^{*})$ is the potential difference between the electrode and the electrolyte (a function of time only since the electrical conductivities are assumed large), and
$U^{*}(c_{s}^{*},c_{e}^{*})$ is the open circuit potential (OCP), an empirical function depending on $c_{e}^{*}$ and the local surface concentration $c_{s}^{*}(t;R)=c^{*}|_{r^{*}=R^{*}}$,
typically found by fitting to experimental measurements relative to
a Li/Li$^{+}$ metal electrode\textemdash a half cell as modelled
here. (As $c_{e}^{*}$ is assumed constant, the dependence of $U^{*}$ on $c_{e}^{*}$ will be subsequently suppressed.) 

Finally, conservation of charge implies that the total lithium flux
out of the electrode is proportional to the applied circuit current
density, $I_{\text{app}}^{*}$. Since the surface area per unit electrode
volume for the particles of radius $R^{*}$ is $a^{*}(R^{*})$, given
by (\ref{eq:area_density}),  charge conservation gives
\begin{equation}
L^{*}\int_{0}^{\infty}a^{*}(R^{*})G^{*}(c_{s}^{*},\Delta\phi^{*})\,\mathrm{d}R^{*}=-\frac{I_{\text{app}}^{*}}{F^{*}},\label{eq:dimensional_I_app-1}
\end{equation}
where $I_{\text{app}}^{*}\gtrless0$ corresponds to a discharge/charge
current density for a cathode half cell (the opposite for an anode
half cell), respectively, and $L^{*}$ is the through-cell thickness
of the electrode (excluding separator and current collector; see Fig.
\ref{fig:MPM-schematic}$(a)$). If the current density $I_{\text{app}}^{*}$
is prescribed, then the unknown variables are $c^{*}(r^{*},t^{*};R^{*})$
for each particle size $R^{*}$, and the potential difference $\Delta\phi^{*}(t^{*})$,
which is common to all particle sizes. If we choose the potential
in the Li metal as our reference, then uniformity of the electrolyte
potential gives $\phi_{e}^{*}\equiv0$ and the half-cell potential
is then simply
$V_{\text{half-cell}}^{*}(t^{*})=\Delta\phi^{*}(t^{*})=\phi^{*}(t^{*})$.

The system (\ref{eq:dimensional_diffusion_eq})-(\ref{eq:dimensional_Butler_Volmer})
consists of a diffusion equation for each $0<R^{*}<\infty$, all coupled
via the integral condition (\ref{eq:dimensional_I_app-1}).

\subsection{Scaling and nondimensional equations}

We first identify the key 
time\-scales present in the problem. As our temporal scaling we use the
 characteristic discharge
timescale for the cell,
defined as
\begin{equation}
\tau_{\text{d}}^{*}=\frac{F^{*}c_{\text{max}}^{*}L^{*}v_{\mathrm{total}}}{I_{\text{typ}}^{*}}.\label{eq:discharge_timescale}
\end{equation}
This corresponds to the time taken for a current density $I_{\mathrm{typ}}^{*}$
to completely discharge the electrode from its theoretical maximum
capacity. There is a diffusion and a reaction timescale,
\begin{equation}
\tau_{\text{diff}}^{*}  =\frac{(R_{\text{typ}}^{*})^{2}}{D^{*}}, \qquad
\tau_{\text{reac}}^{*}  =\frac{F^{*}}{k^{*}a_{\text{typ}}^{*}(c_{e}^{*})^{1/2}},
\end{equation}
respectively, where $R_{\text{typ}}^{*}$ and $a_{\text{typ}}^{*}$ are typical
scales for the particle radii and surface area per unit electrode
volume. We choose $a_{\mathrm{typ}}^{*}$ to be the surface area if
all particles were of the typical radius, i.e., $a_{\text{typ}}^{*}=3v_{\mathrm{total}}/R_{\text{typ}}^{*}$.
We will make different choices for $R_{\text{typ}}^{*}$ depending
on whether the particle size distribution is unimodal or bimodal.
For a unimodal particle size distribution, it is natural to take $R_{\text{typ}}^{*}=\overline{R}_{n}^{*}$,
however, for a bimodal distribution it is more convenient to use a
typical radius based on the average of just one of the modes or components\textemdash see
\S\ref{sec:Bimodal-Particle-Size-Distributi}. 

Particle radii and radial coordinates are scaled with $R_{\text{typ}}^{*}$,
concentrations with the maximum concentration $c_{\text{max}}^{*}$,
time with the discharge timescale, potentials with $\Phi^{*}=1$ V,
and the current density with $I_{\mathrm{typ}}^{*}$ (this allows
the consideration of nonconstant applied current densities).
The Butler\textendash Volmer reaction rate is scaled based on the
typical lithium flux required to sustain the current density $I_{\mathrm{typ}}^{*}$.
By scaling the total surface area $a_{\mathrm{total}}^{*}$
with $a_{\text{typ}}^{*}$, natural scalings for all PSD quantities follow from their definitions (\ref{eq:A_and_V})-(\ref{eq:variance_i_star})
and appropriate powers of $R_{\mathrm{typ}}^{*}$. The resulting
scalings are

\begin{align*}
r^{*} & =R_{\text{typ}}^{*}r, & (c^{*},c_{s}^{*},c_{\text{init}}^{*}) & =c_{\text{max}}^{*}(c,c_{s},c_{\text{init}}), & t^{*} & =\tau_{\text{d}}^{*}t,\\
I_{\text{app}}^{*} & =I_{\text{typ}}^{*}I, & (\eta^{*},\Delta\phi^{*},U^{*}) & =\Phi^{*}(\eta,\Delta\phi,U), & G^{*} & =\frac{I_{\text{typ}}^{*}}{F^{*}L^{*}a_{\mathrm{typ}}^{*}}G,\\[-11mm]
\end{align*}
\begin{align*}
R^{*} & =R_{\text{typ}}^{*}R, & n_{\text{total}}^{*} & =\frac{a_{\text{typ}}^{*}}{R_{\text{typ}}^{*2}}n_{\text{total}}, & a_{\text{total}}^{*} & =a_{\text{typ}}^{*}a_{\text{total}},&
n^{*} & =\frac{a_{\text{typ}}^{*}}{R_{\text{typ}}^{*3}}n,\\
 a^{*} & =\frac{a_{\text{typ}}^{*}}{R_{\text{typ}}^{*}}a, & v^{*} & =a_{\text{typ}}^{*}v,&
f_{i}^{*} & =\frac{1}{R_{\text{typ}}^{*}}f_{i}, & m_{i,j}^{*} & =R_{\text{typ}}^{*j}m_{i,j},
\end{align*}
which transform (\ref{eq:dimensional_diffusion_eq})-(\ref{eq:dimensional_I_app-1})
into the nondimensional problem
\begin{align}
\frac{\partial c}{\partial t} & =\gamma\frac{1}{r^{2}}\frac{\partial}{\partial r}\left(r^{2}\frac{\partial c}{\partial r}\right),\qquad\text{for }0<r<R,\label{eq:diffusion_eq}\\
c & =c_{\text{init}},\qquad\text{at }t=0,
\end{align}
with regularity at $r=0$ and boundary condition
\begin{align}
%\frac{\partial c}{\partial r} & =0,\qquad\text{at }r=0,\label{eq:BC_centre}\\
-\gamma\frac{\partial c}{\partial r} & =\frac{1}{3}G(c,\Delta\phi)\qquad\text{at }r=R,\label{eq:BC_surface}
\end{align}
which are coupled via the $\Delta \phi$-dependence of their intercalation rates
\begin{align}
  G(c_{s},\Delta\phi) & =g(c_{s})\sinh\left\{ \frac{\lambda}{2}\left[\Delta\phi-U(c_{s})\right]\right\} ,\label{eq:G_PSD-1}
  \qquad \displaystyle g(c_{s})  =kc_{s}^{1/2}\left(1-c_{s}\right)^{1/2},
\end{align}
which must satisfy the total flux constraint
\begin{equation}
\int_{0}^{\infty}a(R)G(c_{s},\Delta\phi(t))\mathrm{d}R=-I(t).\label{eq:Q_PSD-2}
\end{equation}
We remark that the factor of $1/3$ arises in boundary condition (\ref{eq:BC_surface})
(and in subsequent equations) because we consider spherical particles.
For different particle shapes this numerical factor would
be different.

All PSD relations (\ref{eq:A_and_V})-(\ref{eq:variance_i_star})
remain the same but with the stars now removed and the total active
volume fraction, $v_{\mathrm{total}}$, replaced by $1/3$. In particular,
\begin{comment}
  \begin{align}
\frac{1}{3} & =\int_{0}^{\infty}v(R)\mathrm{d}R,\label{eq:v_total}\\
f_{v}(R) & =3v(R),\\
n_{\mathrm{total}} & =\frac{1}{4\pi\int_{0}^{\infty}R^{3}f_{n}(R)\mathrm{d}R}=\frac{1}{4\pi m_{n,3}},\label{eq:n_total}\\
a_{\mathrm{total}} & =\frac{\int_{0}^{\infty}R^{2}f_{n}(R)\mathrm{d}R}{\int_{0}^{\infty}R^{3}f_{n}(R)\mathrm{d}R}=\frac{m_{n,2}}{m_{n,3}}=\frac{1}{\bar{R}_{a}}.\label{eq:a_total}
\end{align}
\end{comment}
\[\int_{0}^{\infty}v(R)\mathrm{d}R= \frac{1}{3}, \qquad
f_{v}(R)  =3v(R), \]
\[
n_{\mathrm{total}}  =\frac{1}{4\pi\int_{0}^{\infty}R^{3}f_{n}(R)\mathrm{d}R}=\frac{1}{4\pi m_{n,3}},
\qquad
a_{\mathrm{total}}  =\frac{\int_{0}^{\infty}R^{2}f_{n}(R)\mathrm{d}R}{\int_{0}^{\infty}R^{3}f_{n}(R)\mathrm{d}R}=\frac{m_{n,2}}{m_{n,3}}=\frac{1}{\bar{R}_{a}}.
\]
The nondimensional parameter groups introduced are 
\begin{align}
\gamma & =\frac{\tau_{\text{d}}^{*}}{\tau_{\text{diff}}^{*}}, & k & =\frac{\tau_{\text{d}}^{*}}{\tau_{\text{reac}}^{*}}, & \lambda & =\frac{F^{*}\Phi^{*}}{R_{g}^{*}T^{*}}.
\end{align}
The parameter $\lambda\approx 38.92$ is the ratio of the typical half cell voltage
to the thermal voltage and is constant for the isothermal problem we consider.
 The
parameters $\gamma$ and $k$ are ratios of the discharge timescale
to the diffusion and reaction timescales, and they act as nondimensional
diffusion and reaction coefficients in (\ref{eq:diffusion_eq})-(\ref{eq:Q_PSD-2}).
Since both $\gamma$ and $k$ depend on the discharge time and thus
the applied current, it can be useful to express them explicitly in
terms of the so-called C-rate. If $\mathrm{C}^{*}$ is the current
density that discharges the half-cell in 1 hour, then the C-rate is
defined as $\mathcal{C}=I_{\mathrm{typ}}^{*}/\mathrm{C}^{*}$. Substituting
$I_{\text{typ}}^{*}=\mathcal{C}\text{C}^{*}$ into (\ref{eq:discharge_timescale}), we can write
$\gamma=\hat{\gamma}/\mathcal{C}$ and $k=\hat{k}/\mathcal{C}$ where $\hat{\gamma}$ and $\hat{k}$ depend only on the material parameters and geometry.

\subsection{Fast diffusion in the electrode particles}
\label{sec:fastdiffusionmodel}
The model may be simplified further if diffusion within the electrode
particles is also fast in comparison to the discharge timescale, i.e.
$\gamma=\tau_{\mathrm{d}}^{*}/\tau_{\mathrm{diff}}^{*}\gg1$. In this
case the MPM (\ref{eq:diffusion_eq})-(\ref{eq:Q_PSD-2}) reduces
to a system of ODEs. It is straightforward to show in the limit $\gamma\to\infty$
that $c$ becomes independent of $r$. Then (\ref{eq:diffusion_eq})-(\ref{eq:Q_PSD-2})
becomes the system of mass balances
\begin{align}
\frac{\mathrm{d}c}{\mathrm{d}t} & =\frac{A(R)}{3V(R)}G(c,\Delta\phi),\qquad0<R<\infty,\ 0<t<\infty,\label{eq:MPM_fast_diffusion}\\
c & =c_{\mathrm{init}},\qquad\text{at }t=0,\\
-I(t) & =\int_{0}^{\infty}a(R)G(c,\Delta\phi)\mathrm{d}R,\label{eq:MPM_fast_diffusion_constraint}
\end{align}
where $A(R)$ and $V(R)$ are the surface area and volume of a particle
of radius $R$. For spherical particles considered here, $A/(\text{3}V)=1/R$.
If all the particles are the same shape but can be parametrised by
a single length parameter $R$, then the system (\ref{eq:MPM_fast_diffusion})-(\ref{eq:MPM_fast_diffusion_constraint})
in fact holds for particles of any shape, and $A/(3V)=\xi/R$ with
the numerical constant $\xi$ depending on the shape.

\subsection{Parameter values}

For most of the paper we consider a graphite anode (meso-carbon microbeads
or LiC$_{6}$) in an electrolyte consisting of LiPF$_{6}$ in the
solvent EC:DMC of ratio 1:1 by volume. The parameters (taken from \cite{Marquis2019}) and resulting
nondimensional parameters used
%(who in turn adapted them from \cite{MouraFastDFN} and \cite{NewmanDualfoil})
are given in Tables \ref{tab:Dimensional-parameters}
and \ref{tab:Dimensionless-parameters} in appendix \ref{sec:Parameter_values_SM}, respectively.
The functional expression for the open circuit potential $U^{*}(c_{s}^{*},c_e^*)$
is taken from the Dualfoil code of Newman \cite{NewmanDualfoil}.
Later, in \S\ref{sec4.3}, we will also compare
with experimental results on a lithium iron phosphate (LiFePO$_{4}$)
cathode in the same electrolyte.

The shape of the PSD is specified by the radius distribution $f_{n}(R)$.
We model a unimodal distribution as a log-normal (typical for electrode
materials \cite{Meyers2000,Song2013}):
\begin{align}
f_{n}(R) & =\frac{1}{R\sqrt{2\pi\sigma_{LN}^{2}}}\exp\left[-\frac{(\log R-\mu_{LN})^{2}}{2\sigma_{LN}^{2}}\right],\qquad\mu_{LN}\in(-\infty,\infty),\,\sigma_{LN}>0,\label{eq:lognormal}
\end{align}
for which
\begin{align}
\bar{R}_{n} & =\exp(\mu_{LN}+\sigma_{LN}^{2}/2),\\
\sigma_{n}^{2} & =\exp(\sigma_{LN}^{2}-1)\exp(2\mu_{LN}+\sigma_{LN}^{2}),\\
m_{n,j} & =\exp(j\mu_{LN}+j^{2}\sigma_{LN}^{2}/2),\qquad j=1,2,...
\end{align}
This can be described by only two parameters, the mean $\bar{R}_{n}$
and standard deviation $\sigma_{n}$. If $\bar{R}_{n}$ is fixed to
be unity by the nondimensionalisation (i.e., $R_{\mathrm{typ}}^{*}=\bar{R}_{n}^{*}$),
then in  \cite{Buqa2005} fitting a log-normal to the PSDs of different graphite  anode
materials gave standard deviations in
the range $0.15\leq\sigma_{n}\leq1$. If we consider only meso-carbon
microbeads (MCMB) \cite{Buqa2005}, then this range is narrowed to
$0.15\leq\sigma_{n}\leq0.3$. Thus we will consider PSDs in the range
of $\sigma_{n}\leq0.3$ for graphite.

Results for Weibull distributions, also commonly
used for PSDs \cite{Roder2016.}, were found to be
very similar for the values of $\sigma_{n}$ considered in this paper.
The bimodal distributions in \S\ref{sec:Bimodal-Particle-Size-Distributi}
are constructed from two log-normal distributions, one for each mode.

\section{Unimodal Particle-Size Distributions}
\label{sec:Unimodel-PSDs}

In this section we consider PSDs that are unimodal, such as that in
Fig.~\ref{fig:MPM-schematic}$(b)$, and derive and assess several
candidate asymptotic solutions in the limit where the distribution
is narrow. We begin by discussing approximations of the PSD using
particles of a single size, and then seek asymptotic corrections to
a selection of these.

\subsection{Single particle models (SPMs)}
\label{subsec:Single-particle-models}
It is commonplace in porous electrode theory to assume that all the
active electrode particles are of the same size. This reduces the
computational complexity considerably, but the question of which particle
radius best represents the full PSD is rarely considered. The model
(\ref{eq:diffusion_eq})-(\ref{eq:Q_PSD-2})
reduces to the single particle model (SPM) when
$f_{n}(R)=\delta(R-R^{\mathrm{SPM}})$, where $\delta$ is the Dirac $\delta$-function. Then, with $c(r,t,R^{\mathrm{SPM}}) = \cspm(r,t)$,
\begin{align}
\frac{\partial \cspm}{\partial t} & =\gamma\frac{1}{r^{2}}\frac{\partial}{\partial r}\left(r^{2}\frac{\partial \cspm}{\partial r}\right),\qquad\text{for }0<r<R^{\mathrm{SPM}},\label{eq:SPM_diffusion_eq}\\
-\gamma\frac{\partial \cspm}{\partial r} & =\frac{1}{3}G^{\mathrm{SPM}}(\cspm,\Delta\phi^{\mathrm{SPM}})\qquad\text{at }r=R^{\mathrm{SPM}},\label{eq:SPM_surface_BC}
\end{align}
with regularity at $r=0$, $\cspm  = \cspm_{\text{init}}$ at $t=0$, and the surface flux $G^{\mathrm{SPM}}$ given by
\begin{equation}
-I(t)=a_{\mathrm{total}}^{\mathrm{SPM}}G^{\mathrm{SPM}},\label{eq:Q_PSD-2-1}
\end{equation}
while
\beq
n_{\mathrm{total}}^{\mathrm{SPM}}  =\frac{1}{4\pi(R^{\mathrm{SPM}})^{3}},\qquad
a_{\mathrm{total}}^{\mathrm{SPM}}  =\frac{1}{R^{\mathrm{SPM}}}.\label{eq:a_total_SPM}
\eeq
Recall that we choose the number of particles (and hence surface area)
so as to fix the total volume, hence  (\ref{eq:a_total_SPM})
depend on the particle radius.

As the surface flux is known exactly in terms of the current density,
$I(t)$, the solution is very straightforward if one prescribes $I(t)$,
since one can solve (\ref{eq:SPM_diffusion_eq})-(\ref{eq:SPM_surface_BC})
for $\cspm$ first then rearrange (\ref{eq:Q_PSD-2-1}) for the potential:
\begin{equation}
\Delta\phi^{\mathrm{SPM}}=U(c_{s}^{\mathrm{SPM}})-\frac{2}{\lambda}\sinh^{-1}\left[\frac{R^{\mathrm{SPM}}I(t)}{g(c_{s}^{\mathrm{SPM}})}\right].\label{eq:Deltaphi_SPM}
\end{equation}
This is considerably simpler than the full MPM (\ref{eq:diffusion_eq})-(\ref{eq:Q_PSD-2}),
which has a continuum of PDEs coupled via an integral equation, even
for prescribed $I(t)$.

The possible choices of $R^{\mathrm{SPM}}$ to represent the PSD are
numerous and should be motivated by the application. From the theory
of PSD characterisation, particularly that of droplets where many
physical processes may be important \cite{BayvelBook}, most of the
important average radii are encapsulated by the following two parameter
family, in terms of the moments of $f_{n}(R)$:
\begin{align}
R[p,q] & =\left(\frac{\int_{0}^{\infty}R^{p}f_{n}(R)\mathrm{d}R}{\int_{0}^{\infty}R^{q}f_{n}(R)\mathrm{d}R}\right)^{1/(p-q)}=\left(\frac{m_{n,p}}{m_{n,q}}\right)^{1/(p-q)},
\quad p\not=q =0,1,2,\ldots.
\end{align}
In this framework, $R[1,0]$ is simply the mean radius, i.e., the
mean $\bar{R}_{n}$ of the distribution $f_{n}(R)$. The mean $R[2,0]$
corresponds to the radius of a sphere with area equal to the average
area of the PSD. Similarly, $R[3,0]$ is the radius of a sphere with
volume equal to the average volume of the PSD. For means of the form
$R[p,0]$, knowledge of the number of particles (in a representative
sample) is necessary. Thus they are all determined with similar measurement
techniques, e.g., microscopy and image analysis \cite{AllenBook},
where individual particles are able to be counted.
%Hence, of this type of mean, we will consider only $R[1,0]=\bar{R}_{n}$. 

A commonly used radius is $R[3,2]$, known as the Sauter mean radius
\cite{BayvelBook} or surface-area moment mean,
\begin{align}
R[3,2] & =\frac{\int_{0}^{\infty}R^{3}f_{n}(R)\mathrm{d}R}{\int_{0}^{\infty}R^{2}f_{n}(R)\mathrm{d}R}=\frac{3\mathbb{E}_{n}[V]}{\mathbb{E}_{n}[A]}=\frac{3\cdot\frac{1}{3}}{a_{\mathrm{total}}}=\frac{1}{a_{\mathrm{total}}},
\end{align}
which corresponds here to $\bar{R}_{a}$, the mean of the area-weighted
radius distribution $f_{a}(R)$. It is widely used for applications
where the active surface area is relevant, such as catalysis and combustion
of liquid sprays \cite{Wang2013}. Its definition is proportional
to the ratio of the average volume $\mathbb{E}_{n}[V]$ to the average
surface area $\mathbb{E}_{n}[A]$\textemdash it corresponds to the
radius of a particle that has the same surface-area-to-volume ratio
as the PSD \cite{Kowalczuk2016}. As we have already fixed
the total electrode volume, this means it is
the unique radius that gives the same total surface area $a_{\mathrm{total}}$
as the PSD.
% This is confirmed by equating the areas (\ref{eq:a_total_SPM})
%and (\ref{eq:n_total}), which gives a radius $R^{\mathrm{SPM}}=\bar{R}_{a}=R[3,2]$.

Another important radius, $R[4,3]$, is known as the de Brouckere
mean radius \cite{BayvelBook} or volume moment mean,
\begin{align}
R[4,3] & =\frac{\int_{0}^{\infty}R^{4}f_{n}(R)\mathrm{d}R}{\int_{0}^{\infty}R^{3}f_{n}(R)\mathrm{d}R},
\end{align}
and corresponds here to $\bar{R}_{v}$, the mean of the volume-weighted
radius distribution $f_{v}(R)$. It is used when the bulk volume is
considered more relevant, used also for combustion \cite{Wang2013}.
Typically $R[4,3]>R[3,2]$, i.e., $\bar{R}_{v}>\bar{R}_{a}$, and
so an SPM of radius $\bar{R}_{v}$ has a lower surface area than the
PSD (see (\ref{eq:a_total_SPM})). However, $f_{v}(R)$ (and
hence $\bar{R}_{v}$) is measured directly by laser diffraction analysis\textemdash one
of the most common and accurate modern PSD analysis techniques \cite{MalvernWP}\textemdash and
hence has significant practical relevance. Indeed, it is also generated
by any measurement technique that is based on mass, such as sieving
and sedimentation \cite{AllenBook}.

The relative sizes of these different means are shown in Table \ref{tab:mean-radii},
given a mean radius $\bar{R}_{n}=1$ and standard deviation $\sigma_{n}=0.3$.
Other mean radii, e.g. $R[2,1]$, $R[3,1]$, and those based
on percentiles have no particular significance in this context and
we do not consider them here.

\begin{table}
\begin{centering}
\begin{tabular}{lcccccc}
\hline 
Mean radius & $R[1,0]$ & $R[2,0]$ & $R[3,0]$ & $R[3,2]$ & $R[4,3]$ & $R[5,3]$\tabularnewline
\hline 
Value & 1 & 1.044 & 1.091 & 1.188 & 1.295 & 1.352\tabularnewline
\hline 
\end{tabular}
\par\end{centering}
\caption{Sizes of mean radii discussed in \S\ref{subsec:Single-particle-models}
for a log-normal distribution $f_{n}(R)$ with mean $\bar{R}_{n}=R[1,0]=1$
and standard deviation $\sigma_{n}=0.3$.}
\label{tab:mean-radii}
\end{table}

\subsubsection{Equivalent capacity radius}
\label{subsec:Equivalent-capacity-radius}
In the
present context we can identify another mean radius that arises naturally
from charge capacity considerations. As shown by R\"oder et al \cite{Roder2016.},
the spread of a PSD can have a significant effect on the discharge
capacity of the half-cell. We will show that 
\begin{align}
\bar{R}_{c}=R[5,3] & =\left(\frac{\int_{0}^{\infty}R^{5}f_{n}(R)\mathrm{d}R}{\int_{0}^{\infty}R^{3}f_{n}(R)\mathrm{d}R}\right)^{1/2},\label{eq:R_c}
\end{align}
is the radius that, when used in an SPM, will exhibit a similar capacity
to the full PSD if $\gamma$ is  large, which is
usually the case  except at high C-rates (see Table \ref{tab:Dimensionless-parameters}).  We
make the argument for a discharging anode, but a similar argument
holds whether charging or discharging an anode or cathode.

For a constant current discharge of an anode, we have $I=-1$, and
the lithium leaves the electrode particles causing $c(r,t;R)$ to
decrease and the potential $\Delta\phi(t)$ to increase. As the surface
concentration $c_{s}$ of any particle approaches zero, $\Delta\phi\to\infty$,
and since $\Delta\phi$ is the same for all particles, this can only
happen if $c_{s}\to0$ for all particles simultaneously. This occurs
in finite time, say at $t=t_{\mathrm{f}}$, and any lithium still
remaining inside the particles is not accessed. The proportion remaining
is larger for slower lithium diffusion within the particles, larger
particles, or higher C-rates. To estimate the amount of lithium remaining
at final time, we use the approximate solution for $\gamma\gg1$.
Substituting a regular expansion of $c$ in powers of $\gamma^{-1}$
into (\ref{eq:diffusion_eq})-(\ref{eq:BC_surface}) results in
\begin{equation}
c(r,t;R)=c_{0}(t;R)+\frac{1}{\gamma}\frac{\mathrm{d}c_{0}}{\mathrm{d}t}\left(\frac{1}{6}r^{2}+B(t)\right)+\cdots,\qquad\gamma\gg1,\label{eq:c_large_gamma}
\end{equation}
where $c_{0}(t;R)$ is the solution of the fast diffusion problem
(\ref{eq:MPM_fast_diffusion})-(\ref{eq:MPM_fast_diffusion_constraint}),
and $B(t)$ is a nontrivial function of $t$ which will not be needed.
Setting $c=0$ on the surface $r=R$ at time $t=t_{\mathrm{f}}$,
and subtracting from (\ref{eq:c_large_gamma}) at $t=t_{\mathrm{f}}$
gives
\begin{equation}
c(r,t_{\mathrm{f}};R)=\frac{1}{6\gamma}\left.\frac{\mathrm{d}c_{0}}{\mathrm{d}t}\right|_{t_{\mathrm{f}}}\left(r^{2}-R^{2}\right)+\cdots,\qquad\gamma\gg1.
\end{equation}
Integrating this over all the particles and dividing by the maximum
capacity, we find the fraction $\mathcal{F}_{\mathrm{cap}}$ remaining
at $t=t_{\mathrm{f}}$ to be
\begin{equation}
  \mathcal{F}_{\mathrm{cap}}=\frac{\int_{0}^{\infty}n(R)(\int_{0}^{R}4\pi r^{2}c
    %|_{t_{\mathrm{f}}}
    \,\mathrm{d}r)\,\mathrm{d}R}{\int_{0}^{\infty}V(R)n(R)\,\mathrm{d}R}=\frac{1}{15\gamma}\frac{\int_{0}^{\infty}%\left.
    (\mathrm{d}c_{0}/\mathrm{d}t)
    %\right|_{t_{\mathrm{f}}}
    R^{5}f_{n}(R)\,\mathrm{d}R}{\int_{0}^{\infty}R^{3}f_{n}(R)\mathrm{d}R}+\cdots,\qquad\gamma\gg1.\label{eq:fraction_remaining_1}
\end{equation}
Now, $c_{0}$ itself hits zero slightly later at $t_{\mathrm{f},0}=t_{\mathrm{f}}+O(\gamma^{-1})>t_{\mathrm{f}}$,
where we have
\begin{equation}
c_{0}\sim(t-t_{\mathrm{f},0}),\qquad G_{0}\sim R,\qquad\Delta\phi_{0}\sim-\frac{2}{\lambda}\log(t-t_{\mathrm{f},0}),\qquad\text{as }t\to t_{\mathrm{f},0}^{+}.
\end{equation}
Thus, $\left.(\mathrm{d}c_{0}/\mathrm{d}t)\right|_{t_{\mathrm{f}}}=1+O(\gamma^{-1})$,
which substituted into (\ref{eq:fraction_remaining_1}) gives
\begin{equation}
\mathcal{F}_{\mathrm{cap}}=\frac{(R[5,3])^{2}}{15\gamma}+\cdots,\qquad\gamma\gg1.\label{eq:f}
\end{equation}
Repeating the analysis for the SPM is simpler, resulting in the fraction
\begin{equation}
\mathcal{F}_{\mathrm{cap}}^{\mathrm{SPM}}=\frac{(R^{\mathrm{SPM}})^{2}}{15\gamma}+\cdots,\qquad\gamma\gg1.\label{eq:f_SPM}
\end{equation}
Thus we identify the mean radius $\bar{R}_{c}=R[5,3]$ as that which would give the correct capacity remaining when used in an SPM.

For practical purposes, $\bar{R}_{c}$ may be calculated easily from
the volume-weighted distribution via
\begin{equation}
\bar{R}_{c}=(m_{v,2})^{1/2}=(\bar{R}_{v}^{2}+\sigma_{v}^{2})^{1/2}.
\end{equation}
and can be interpreted as the radius that gives the average surface
area when calculated using the volume-weighted distribution $f_{v}$. 

\subsection{Corrections to single particle models for narrow distributions}
\label{subsec:Asymptotic-corrections}
\label{sec3.2}
The SPM, with a given  choice of effective particle radius,
is the leading-order approximation as $\sigma_n \rightarrow 0$, that is, when the unimodal PSD is narrow and most particles are clustered close to  that single radius.
In this section we seek an asymptotic correction to this SPM. 
We present the analysis for the area-weighted mean $\bar{R}_{a}$
as it is the simplest, with the corresponding results for $\bar{R}_{n}$
and $\bar{R}_{v}$ given in the Supplementary Material. We find it more convenient to use $\sigma_a$ as the small parameter here, noting that  $\sigma_a = O(\sigma_n)$ as $\sigma_n \rightarrow 0$.

In terms of the
area-weighted distribution the integral constraint (\ref{eq:Q_PSD-2})   is
\begin{equation}
-I(t)=\frac{1}{\bar{R}_{a}}\int_{0}^{\infty}f_{a}(R)G(c_{s}(t;R),\Delta\phi(t))\mathrm{d}R.\label{eq:Q_PSD-1-1-3}
\end{equation}
In the limit $\sigma_{a}\to0$, $f_{a}$ will become concentrated
around $R=\bar{R}_{a}$ and therefore we Taylor expand the integrand
(but not the distribution $f_{a}$ itself) about this value
\[
G=G_{a}+(R-\bar{R}_{a})\left(\frac{\partial G}{\partial R}\right)_{a}+\frac{1}{2!}(R-\bar{R}_{a})^{2}\left(\frac{\partial^{2}G}{\partial R^{2}}\right)_{a}+O((R-\bar{R}_{a})^{3}),
\]
where the subscript $a$ denotes evaluation at $R=\bar{R}_{a}$.
Substituting into (\ref{eq:Q_PSD-1-1-3}) and assuming $\sigma_{a}$
is sufficiently small so we can integrate term-by-term gives 
\begin{align}
-I(t)= & \frac{1}{\bar{R}_{a}}\left[G_{a}+\frac{1}{2}\left(\frac{\partial^{2}G}{\partial R^{2}}\right)_{a}\sigma_{a}^{2}+O(\sigma_{a}^{3}\tilde{\mu}_{3,a})\right].
\end{align}
Truncating the
Taylor expansion produces an error due to the skewness, which is $O(\sigma_{a}^{3}\tilde{\mu}_{3,a})=O(\sigma_{a}^{4})$
for a log-normal (\ref{eq:lognormal}), but we have not yet formally
expanded in $\sigma_{a}$. We do so now, expanding all variables in powers of $\sigma_{a}^2$ as
\begin{equation}
  G  =G^{(0)}+\sigma_a^{2}G^{(2)}+\cdots,
\end{equation}
etc. At leading order $G_{a}^{(0)}=-\bar{R}_{a}I(t)$, which
gives exactly
the SPM (\ref{eq:SPM_diffusion_eq})-(\ref{eq:Deltaphi_SPM}) with radius
$R=\bar{R}_{a}$ 
as expected. At $O(\sigma_a^2)$ we find
\beq
 G_{a}^{(2)}=-\frac{1}{2}\left(\frac{\partial^{2}G^{(0)}}{\partial R^{2}}\right)_{a},\label{eq:G_2_a-2}
\eeq
so that the correction $G_{a}^{(2)}$ to the boundary
flux for $R=\bar{R}_{a}$ is given in terms of the second derivative
of $G^{(0)}$ with respect to particle radius $R$ evaluated at $R=\bar{R}_{a}$. Since the SPM (\ref{eq:SPM_diffusion_eq})-(\ref{eq:Deltaphi_SPM})
has $R \equiv \bar{R}_{a}$, to evaluate this second derivative
 we need to solve (two) additional problems, which we describe next. We  consider the
cases of fast and non-fast solid diffusion separately.

\subsubsection{\label{subsec:d2GdR2-fast}Calculation of $(\partial^{2}G^{(0)}/\partial R^{2})_{a}$
for fast diffusion}

When solid diffusion within the particles is fast, the governing equations reduce to (\ref{eq:MPM_fast_diffusion})-(\ref{eq:MPM_fast_diffusion_constraint}),
with $R$ appearing as a parameter in the ODEs. In this case,
we may differentiate (\ref{eq:MPM_fast_diffusion}) directly with
respect to $R$, noting that $G(c,\Delta\phi)$ depends on $R$ only
through $c$, to give
\begin{align}
\left[\frac{\mathrm{d}}{\mathrm{d}t}+\frac{1}{R}\frac{\partial G}{\partial c}\right]\left(\frac{\partial c}{\partial R}\right) & =\frac{1}{R^{2}}G,\label{eq:dcdR}\\
\left[\frac{\mathrm{d}}{\mathrm{d}t}+\frac{1}{R}\frac{\partial G}{\partial c}\right]\left(\frac{\partial^{2}c}{\partial R^{2}}\right) & =-\frac{1}{R}\frac{\partial^{2}G}{\partial c^{2}}\left(\frac{\partial c}{\partial R}\right)^{2}+\frac{2}{R^{2}}\frac{\partial G}{\partial c}\frac{\partial c}{\partial R}-\frac{2}{R^{3}}G.\label{eq:d2cd2R}
\end{align}
If we now substitute $R=\bar{R}_{a}$, and the leading order solutions
for $c_{a}^{(0)}$ and $\Delta\phi^{(0)}$ into $G$ and its derivatives,
equations (\ref{eq:dcdR}) and (\ref{eq:d2cd2R}) form two ODEs for
the unknowns $(\partial c^{(0)}/\partial R)_{a}$ and $(\partial^{2}c^{(0)}/\partial R^{2})_{a}$,
from which we can determine $(\partial^{2}G^{(0)}/\partial R^{2})_{a}$
using the chain rule, i.e.
\begin{align}
\frac{\partial^{2}G}{\partial R^{2}} & =\frac{\partial^{2}G}{\partial c^{2}}\left(\frac{\partial c}{\partial R}\right)^{2}+\frac{\partial G}{\partial c}\frac{\partial^{2}c}{\partial R^{2}}.
\end{align}
Expressions for $\partial G/\partial c$
and $\partial^{2}G/\partial c^{2}$ are given in \S\ref{sec:dGdc_and_d2Gdc2}.

\subsubsection{Calculation of $(\partial^{2}G^{(0)}/\partial R^{2})_{a}$ for non-fast
diffusion}

When  diffusion within the particles is not instantaneous
 it is easiest to
approximate $\partial^{2}G/\partial R^{2}$ by solving (\ref{eq:SPM_diffusion_eq})-(\ref{eq:SPM_surface_BC})
for radii either side and close to the mean, $R=\bar{R}_{a}\pm\Delta R$,
given the leading-order potential $\Delta\phi^{(0)}$. From the solutions
$c_{a,\pm}^{(0)}$ for these neighbouring radii, we can then approximate
the derivative using finite differences, e.g.
\begin{align}
\left(\frac{\partial^{2}G^{(0)}}{\partial R^{2}}\right)_{a} & =\frac{G(c_{a,+,s}^{(0)},\Delta\phi^{(0)})-2G_{a}^{(0)}+G(c_{a,-,s}^{(0)},\Delta\phi^{(0)})}{\Delta R}+O(\Delta R^{2}).\label{eq:d2Gd2R_finite_difference}
\end{align}
This approach is applicable to any form of lithium transport within
the particles since (\ref{eq:d2Gd2R_finite_difference}) applies only
on the surface and has the advantage that the auxiliary problems are
also single particle problems, but with a different particle radius.
It is preferable to a direct boundary perturbation analysis for $R$
close to $\bar{R}_{a}$, which involves significant algebra (as one
must proceed to second order) and computational difficulties in capturing
the behaviour at the particle surface, where spatial gradients become
successively larger at higher orders.

\subsubsection{Corrected potential}

Given $G_{a}^{(2)}$ from (\ref{eq:G_2_a-2}), the problem for the
correction to the concentration $c_{a}^{(2)}$ at the radius $R=\bar{R}_{a}$
is similar to that at leading order, but with a vanishing initial
condition,
\begin{align}
\frac{\partial c_{a}^{(2)}}{\partial t} & =\frac{\gamma}{r^{2}}\frac{\partial}{\partial r}\left[r^{2}\frac{\partial c_{a}^{(2)}}{\partial r}\right], & 0<r & <\bar{R}_{a},0<t\label{eq:PSD_1-1}\\
-\gamma\frac{\partial c_{a}^{(2)}}{\partial r} & =\frac{1}{3}G_{a}^{(2)}, & \text{at }r & =\bar{R}_{a},\label{eq:PSD_3-1}\\
c_{a}^{(2)} & =0, & \text{at }t & =0,
\end{align}
with regularity ar $r=0$.
The potential can then be updated analytically using (\ref{eq:G_PSD-1}) and substituting $c_{a,s}=c_{a,s}^{(0)}+\sigma_{a}^{2}c_{a,s}^{(2)}+O(\sigma_{a}^{4})$ and $G_{a}=G_{a}^{(0)}+\sigma_{a}^{2}G_{a}^{(2)}+O(\sigma_{a}^{4})$, giving
\begin{equation}
\Delta\phi=U(c_{a,s}^{(0)}+\sigma_{a}^{2}c_{a,s}^{(2)})+\frac{2}{\lambda}\sinh^{-1}\left[\frac{G_{a}^{(0)}+\sigma_{a}^{2}G_{a}^{(2)}}{g(c_{a,s}^{(0)}+\sigma_{a}^{2}c_{a,s}^{(2)})}\right]+O(\sigma_{a}^{4}).\label{eq:Deltaphi_SPM-1}
\end{equation}
We could Taylor expand this for $\sigma_{a}\ll1$ but, since it is already explicit, we  keep it in unexpanded form to avoid
 magnified errors near electrode depletion or saturation, where $U$
is singular.

\subsection{Results}
\label{subsec:Unimodal-Results}
To study heterogeneity due to the PSD, and also assess the validity
of the single particle models and their asymptotic corrections, we
consider a constant current discharge of a graphite anode half-cell,
with parameters given in Tables \ref{tab:Dimensional-parameters}
and \ref{tab:Dimensionless-parameters}. During discharge, lithium
deintercalates from the electrode particles, which begin at a uniform
lithium concentration of $c_{\mathrm{init}}=0.8$, and the half-cell
potential  $\Delta\phi^{*}$ increases
until a cut-off value of 0.6 V. The mean radius $\bar{R}_{n}$ is fixed
to be one, and the spread of the PSD is then controlled via the single
parameter $\sigma_{n}$.

\subsubsection{Numerical methods}

The full MPM model (\ref{eq:diffusion_eq})-(\ref{eq:Q_PSD-2}), each
SPM model (\ref{eq:SPM_diffusion_eq})-(\ref{eq:Deltaphi_SPM}) and
their corrections in \S\ref{subsec:Asymptotic-corrections}
were solved numerically using a finite-volume discretisation within
each particle (with 30 volumes being sufficient) and an adaptive explicit
ODE solver in MATLAB for time integration. The solver \emph{ode15s}
was used due to the presence of the algebraic equation resulting from
the integral constraint (\ref{eq:Q_PSD-2}), which we discretise assuming
a finite number of particle sizes, $N$, equispaced between $R=0$
and $R=\bar{R}_{n}+10\sigma_{n}$ (excluding $R=0$). Then $N$ was
increased until convergence, with $N=75$ sufficient for graphical
accuracy. When diffusion is fast and the model
reduces to ODEs,
the computations are simpler, and $N=300$ was taken for the results
presented here.

\subsubsection{Fast diffusion}
\label{subsec:Results-Fast-diffusion}
\begin{figure}
\begin{centering}
\includegraphics[width=0.6\textwidth]{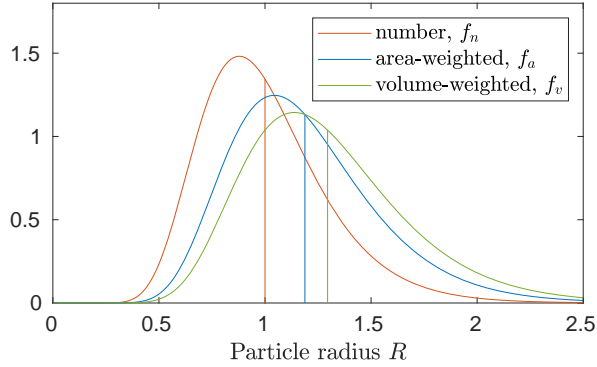}
\par\end{centering}
\caption{Log-normal particle size (number) distribution $f_{n}(R)$ with $\bar{R}_{n}=1$
and $\sigma_{n}=0.3$, and corresponding area-weighted- and volume-weighted-distributions
$f_{a}$,$f_{v}$. Vertical lines indicate the location of their means,
$\bar{R}_{n}=1$, $\bar{R}_{a}=1.188$, $\bar{R}_{v}=1.295$.}

\label{distributions}
\end{figure}

\begin{figure}
\begin{centering}
\includegraphics[width=1\textwidth]{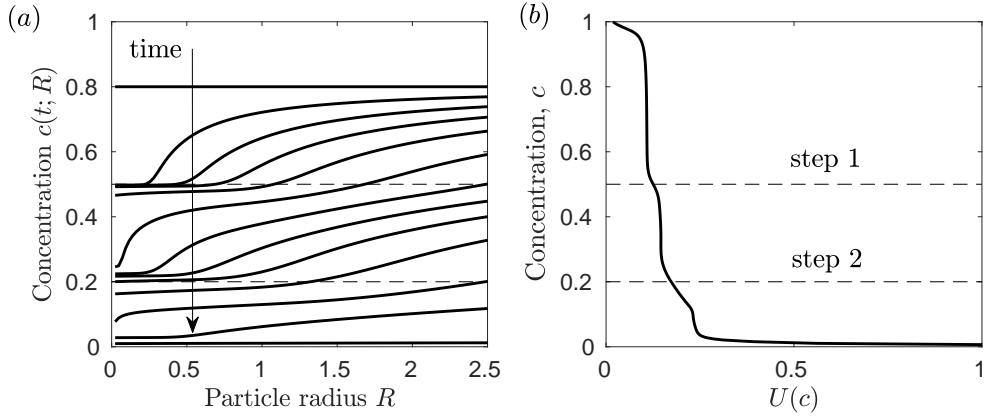}
\par\end{centering}
\caption{A 1C discharge of a graphite anode (fast diffusion $\gamma=\infty$),
for PSD in Fig.~\ref{distributions}: $(a)$
concentrations at selected times and $(b)$ OCP of graphite, $U(c)$.
The concentrations corresponding approximately to the ``steps''
of $U(c)$ are indicated with dashed lines.}

\label{fast_diffusion_c_vs_R}
\end{figure}

First we consider the  case (\ref{eq:MPM_fast_diffusion})-(\ref{eq:MPM_fast_diffusion_constraint}) where the diffusion in the particles
is fast.
A plot of the typical evolution of the concentrations $c(t;R)$ throughout
the discharge is shown in Fig.~\ref{fast_diffusion_c_vs_R}, for a
C-rate $\mathcal{C}=1$ and log-normal PSD with
standard deviation $\sigma_{n}=0.3$ (shown in Fig.~\ref{distributions}). The behaviour shown in Fig.
\ref{fast_diffusion_c_vs_R}$(a)$ can be understood with reference
to the OCP for graphite, $U(c)$, which has several steps\textemdash quick
transitions between plateaus (see Fig.~\ref{fast_diffusion_c_vs_R}$(b)$).
Smaller particles deplete faster due to their larger surface-area-to-volume
ratio,  until their lithium concentration reaches
a steeper section of the OCP. Their reaction resistance increases
and deintercalation shifts to the larger particles. When the lithium
concentration in all particles is again comparable, the smaller particles
deplete more quickly again and the process repeats. Thus we see in \ref{fast_diffusion_c_vs_R}$(a)$
an irregular and staggered depletion of smaller
particles, with larger particles depleting more smoothly.

\begin{figure}
\begin{centering}
\includegraphics[width=0.47\textwidth]{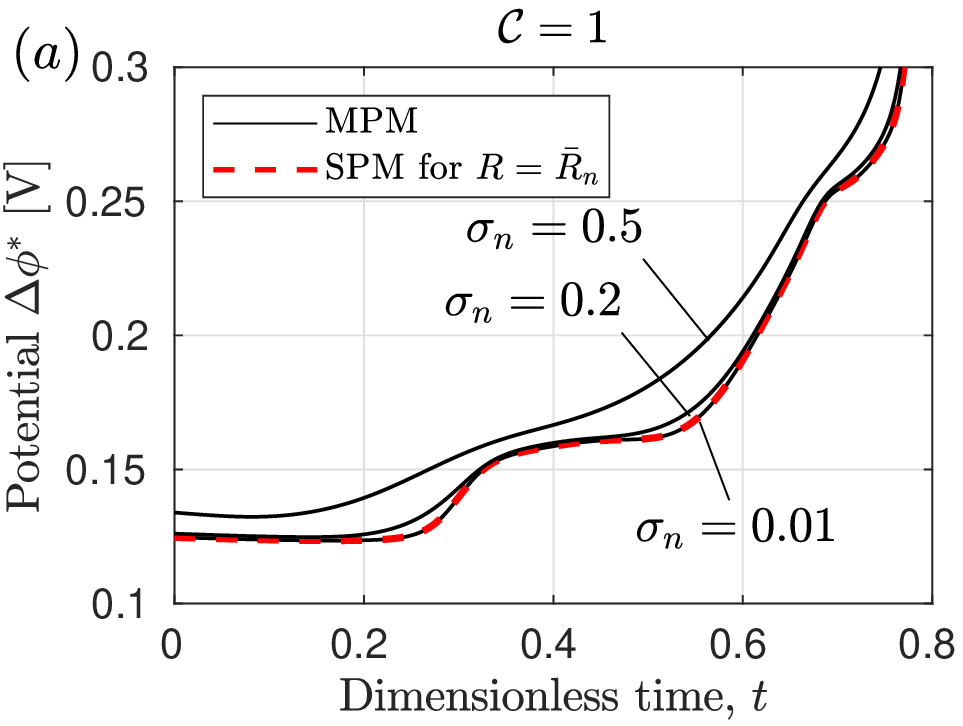}\includegraphics[width=0.47\textwidth]{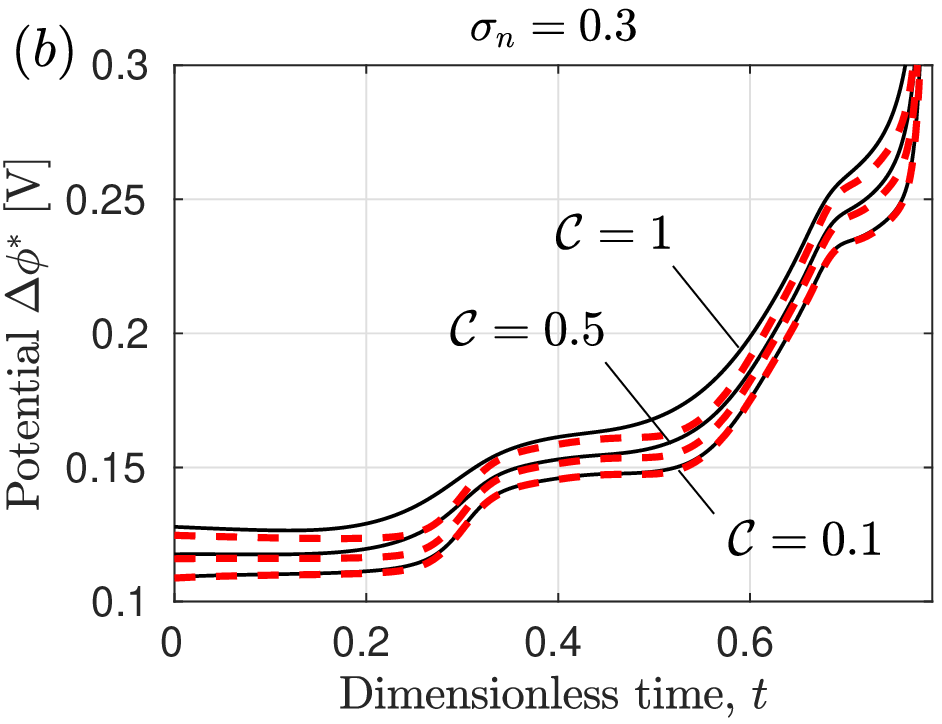}
\par\end{centering}
\caption{Cell potential throughout a discharge (fast diffusion $\gamma=\infty$),
showing MPM and SPM (at $\bar{R}_{n}$) results for: $(a)$ various
$\sigma_{n}$ with C-rate $\mathcal{C}=1$; $(b)$ various $\mathcal{C}$
with $\sigma_{n}=0.3$.}

\label{fast_diffusion_phi_vs_t}
\end{figure}

The consequences for the cell potential are shown in Fig.~\ref{fast_diffusion_phi_vs_t}.
As $\sigma_{n}$ is increased, the increasingly nonuniform discharge
results in both an increase and a smoothing of the potential (Fig.~\ref{fast_diffusion_phi_vs_t}$(a)$).
This effect was not observed by R\"oder et al. \cite{Roder2016.},
as they used an analytical (logarithmic) expression for $U(c)$ for
graphite where the ``steps'' seen in the empirical $U(c)$ are absent.
Furthermore, for a given $\sigma_{n}>0$, the smoothing near the steps
is enhanced as the  C-rate $\mathcal{C}$ increases, as seen  in Fig.
\ref{fast_diffusion_phi_vs_t}$(b)$. This smoothing effect cannot be captured by an SPM. Finally we note that the effect
of PSD spread  on the discharge capacity of the electrode is negligible (not shown), 
since when diffusion is fast all the intercalated lithium can be accessed.

\begin{figure}
\begin{centering}
\includegraphics[width=0.47\textwidth]{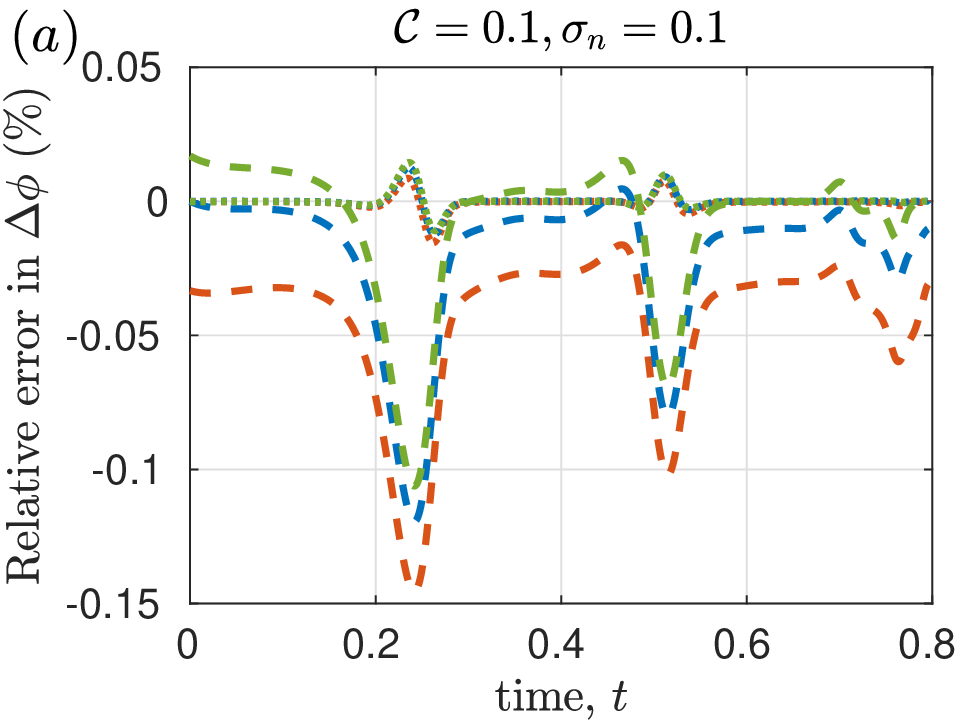}\includegraphics[width=0.47\textwidth]{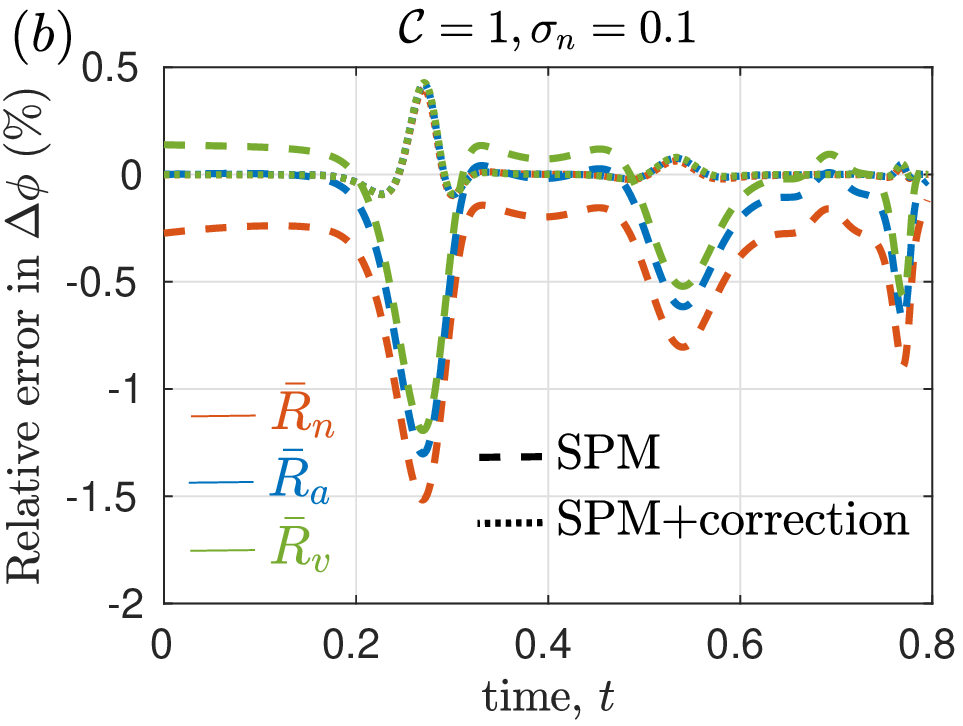}
\par\end{centering}
\begin{centering}
\includegraphics[width=0.47\textwidth]{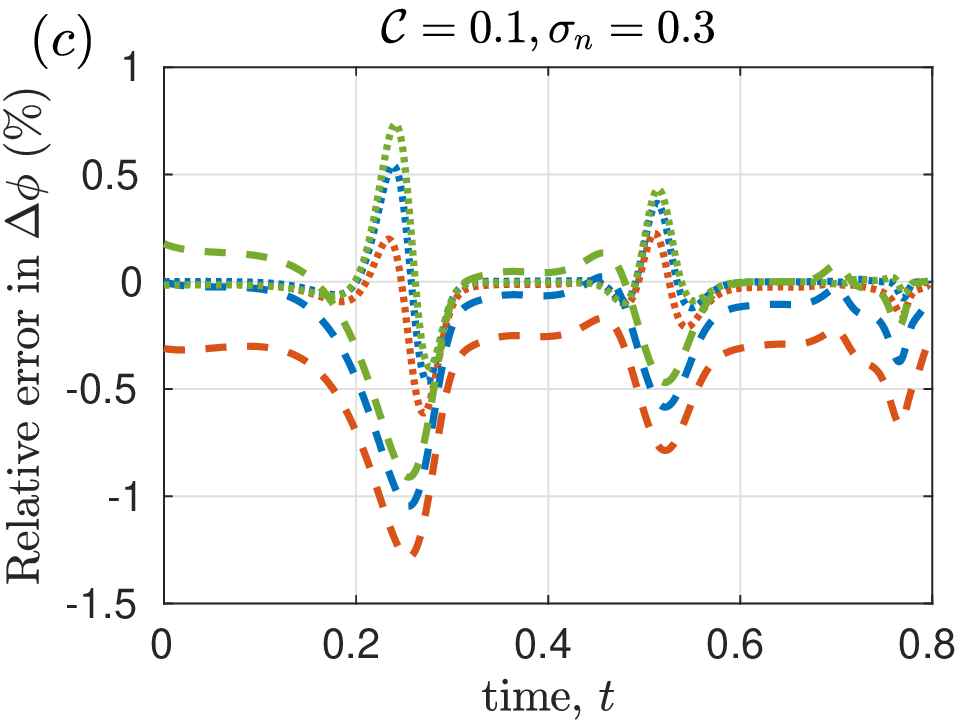}\includegraphics[width=0.47\textwidth]{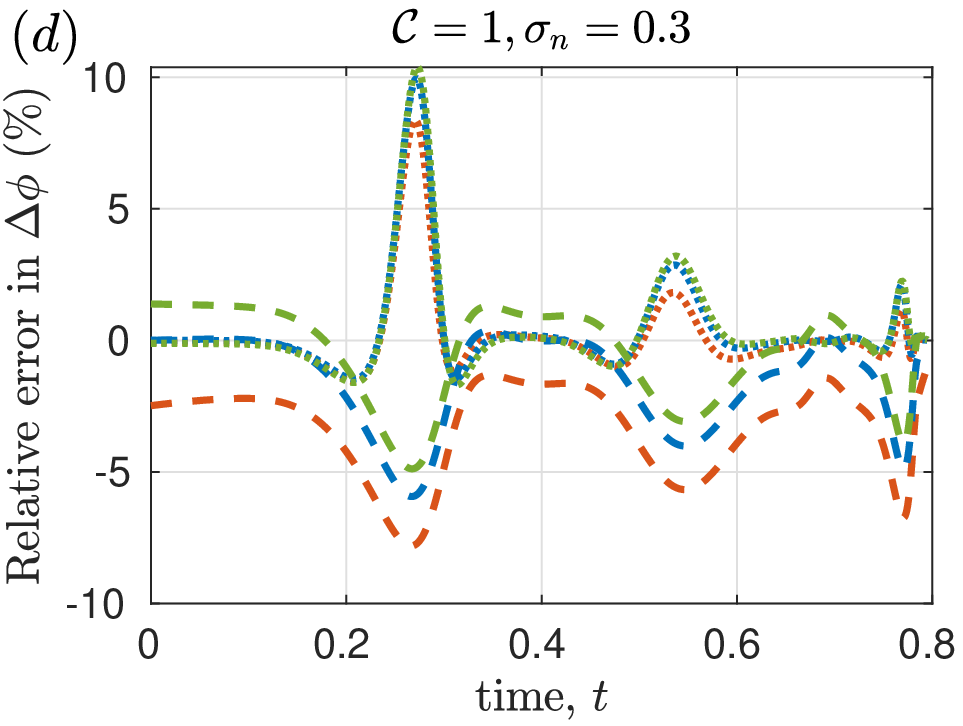}
\par\end{centering}
\caption{Signed error in cell potential relative to MPM throughout a discharge
(fast diffusion $\gamma=\infty$), of SPM (dashed lines) and SPM corrected
for narrow distributions (dotted lines). Results for the mean radius
$\bar{R}_{n}$ (red), area-weighted mean radius $\bar{R}_{a}$ (blue),
and volume-weighted mean radius $\bar{R}_{v}$ (green). Different
panels are for $\sigma_{n}=0.1,0.3$ and C-rate $\mathcal{C}=0.1,1$,
indicated.}

\label{fast_diffusion_error_vs_t}
\end{figure}

To assess the extent to which the SPMs and their corrections for narrow
distributions can approximate the aforementioned behaviour, we plot
the error of each relative to the MPM throughout the discharge, shown
in Fig.~\ref{fast_diffusion_error_vs_t}, for different $\sigma_{n}$
and $\mathcal{C}$. Each SPM (dashed lines) exhibits a significant
negative error (thus underestimate) around $t=0.25$ and $0.55$,
where the concentration in the particles passes the steps in $U(c)$,
because the SPMs cannot capture the smoothing effect. Away from the
steps, where $U(c)$ is relatively flat, the SPMs at $\bar{R}_{n}$
and $\bar{R}_{v}$ under- and overestimate the potential, with the
SPM at $\bar{R}_{a}$ being the most accurate. As described in \S\ref{subsec:Single-particle-models}, this is because $\bar{R}_{a}$
is the unique radius that gives the same surface area as the
PSD, and hence the same reaction resistance if the concentration
in every particle is the same, which is true initially and after
every step in $U(c)$.

\begin{figure}
\begin{centering}
\includegraphics[width=0.55\textwidth]{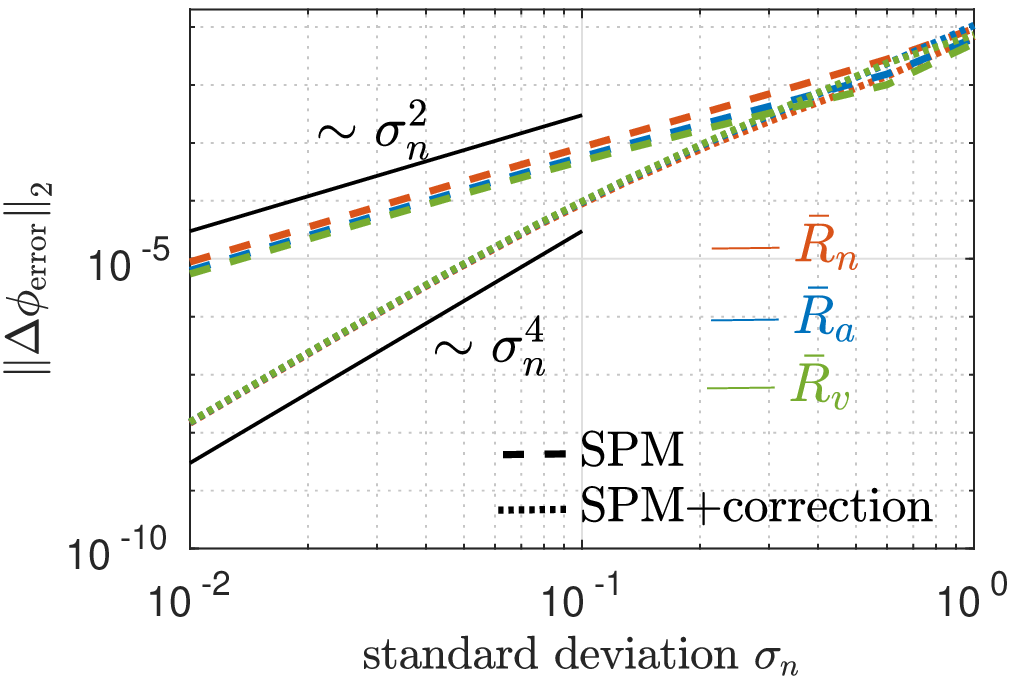}
\par\end{centering}
\caption{Global $L^{2}$-error, $||\Delta\phi_{\mathrm{error}}||_{2}=\left(\int\Delta\phi_{\mathrm{error}}^{2}\mathrm{d}t\right)^{1/2},$
in cell potential relative to MPM throughout a discharge (fast diffusion
$\gamma=\infty$) of SPMs and SPMs corrected for narrow
distributions, $\sigma_{n}\ll1$. See legend and caption of Fig.~\ref{fast_diffusion_error_vs_t}. C-rate
is $\mathcal{C}=1$.}

\label{fast_diffusion_L2_error}
\end{figure}

The corrected SPMs for narrow distributions (dotted lines in Fig.
\ref{fast_diffusion_error_vs_t}) show a significant improvement in
accuracy over the SPMs away from the steps of $U(c)$, for any choice
of effective radius. However, as $\sigma_{n}$ or $\mathcal{C}$ is increased,
the corrected SPMs begin to overcorrect for the smoothing of the steps
and the approximation breaks down locally. This demonstrates the difficulty
in capturing the smoothing effect using the dynamics only close to
the mean radius. In particular, approximating the concentrations $c$
and surface fluxes $G$ with a Taylor expansion in $R$ about a mean
quickly fails to predict the behaviour of particles across the spread
of the PSD as $\sigma_{n}$ increases. Despite this local error, the
global error, e.g. the L$^{2}$-norm of the error over the entire
discharge, shown in Fig.~\ref{fast_diffusion_L2_error}, behaves as
expected. The L$^{2}$-error is $O(\sigma_{n}^{2})$ for the SPMs
and $O(\sigma_{n}^{4})$ for their corrections, with a global improvement
over the SPMs for $\sigma_{n}\leq0.3$, a physically relevant range.

\subsubsection{Non-fast diffusion}
\label{subsec:Finite-diffusion-speed}
\begin{figure}[htbp]
\begin{centering}
\includegraphics[width=0.65\textwidth]{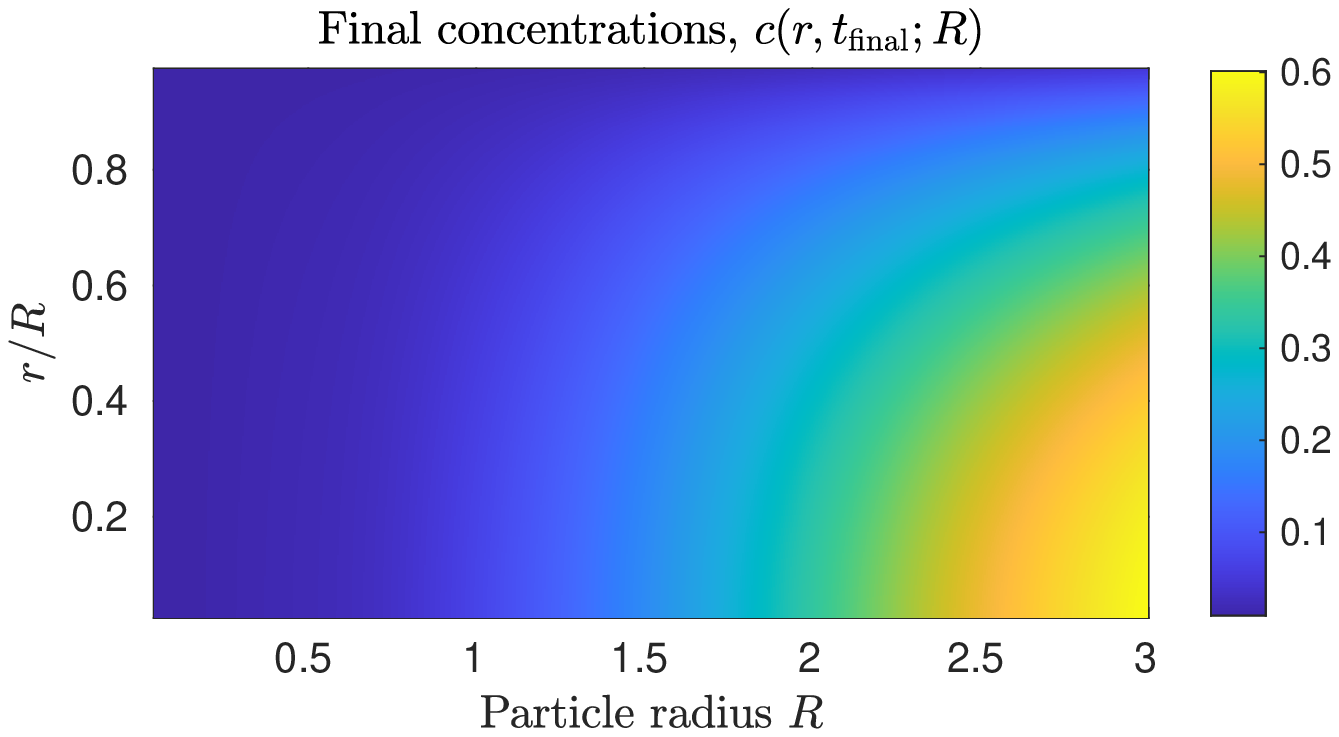}
\par\end{centering}
\caption{Nondimensional  distribution of lithium throughout the
particles at the end of discharge (at voltage cut-off),
for a log-normal distribution (shown in Fig.~\ref{fast_diffusion_c_vs_R}) with
mean $\bar{R}_{n}=1$ and standard deviation $\sigma_{n}=0.3$, and  C-rate
$\mathcal{C}=1$.}

\label{final_concentrations}
\end{figure}

We now consider the full MPM problem  (\ref{eq:diffusion_eq})-(\ref{eq:Q_PSD-2}), accounting for diffusion in the particles.
In this case, as well as the effects described in \S\ref{subsec:Results-Fast-diffusion},
there is an additional, more significant, effect on the usable capacity,
as observed by R\"oder et al. \cite{Roder2016.} for a similar model.
The usable capacity is calculated by Coulomb counting, i.e. integrating
the applied current density in time until voltage cut-off. The amount
discharged by any particular time is the depth of discharge, expressed
here as a percentage of the theoretical maximum, $F^{*}c_{\mathrm{init}}^{*}L^{*}v_{\mathrm{total}}$.
As discussed in \S\ref{subsec:Equivalent-capacity-radius},
at the end of discharge, the surface concentration $c_{s}$ of each
particle approaches zero, and the finite diffusivity results in
lithium remaining in the particle cores, unaccessed. Fig.~\ref{final_concentrations}
shows a typical concentration distribution at voltage cut-off, with
higher concentrations remaining in larger particles. 

\begin{figure}[htbp]
\begin{centering}
\includegraphics[width=0.46\textwidth]{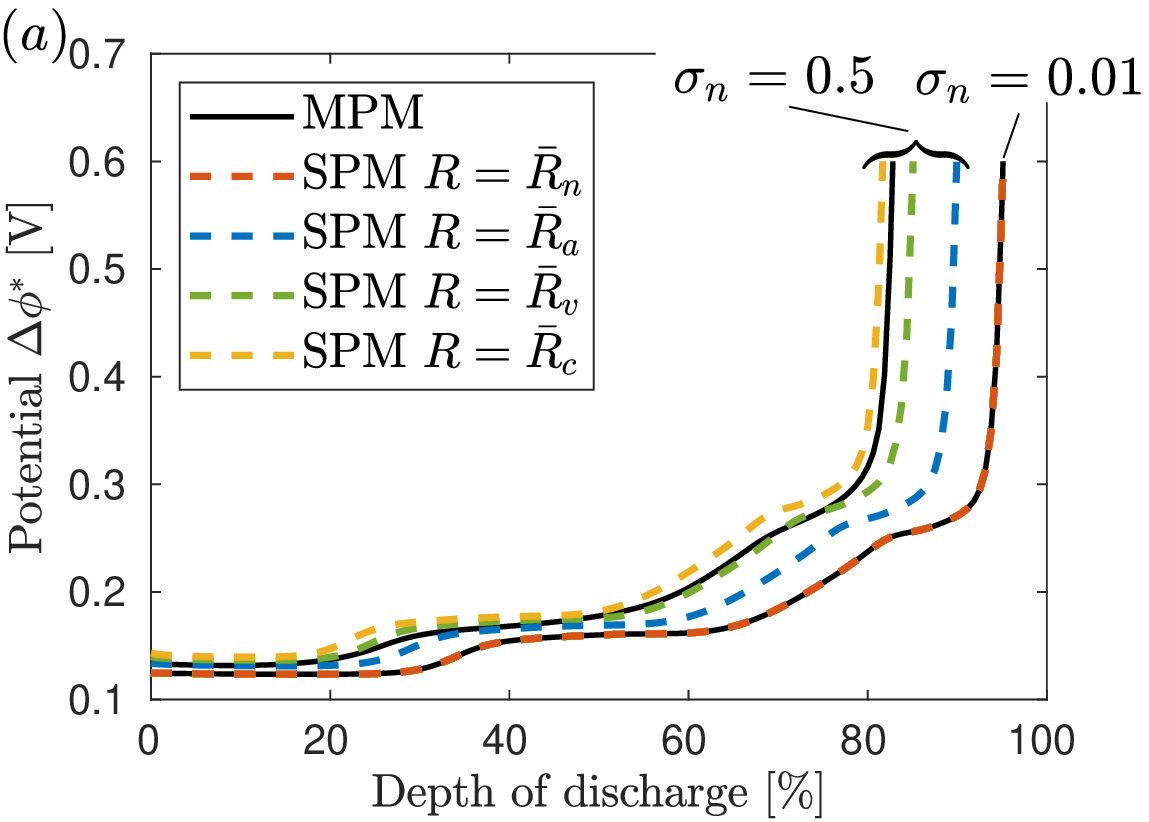}~~~~\includegraphics[width=0.44\textwidth]{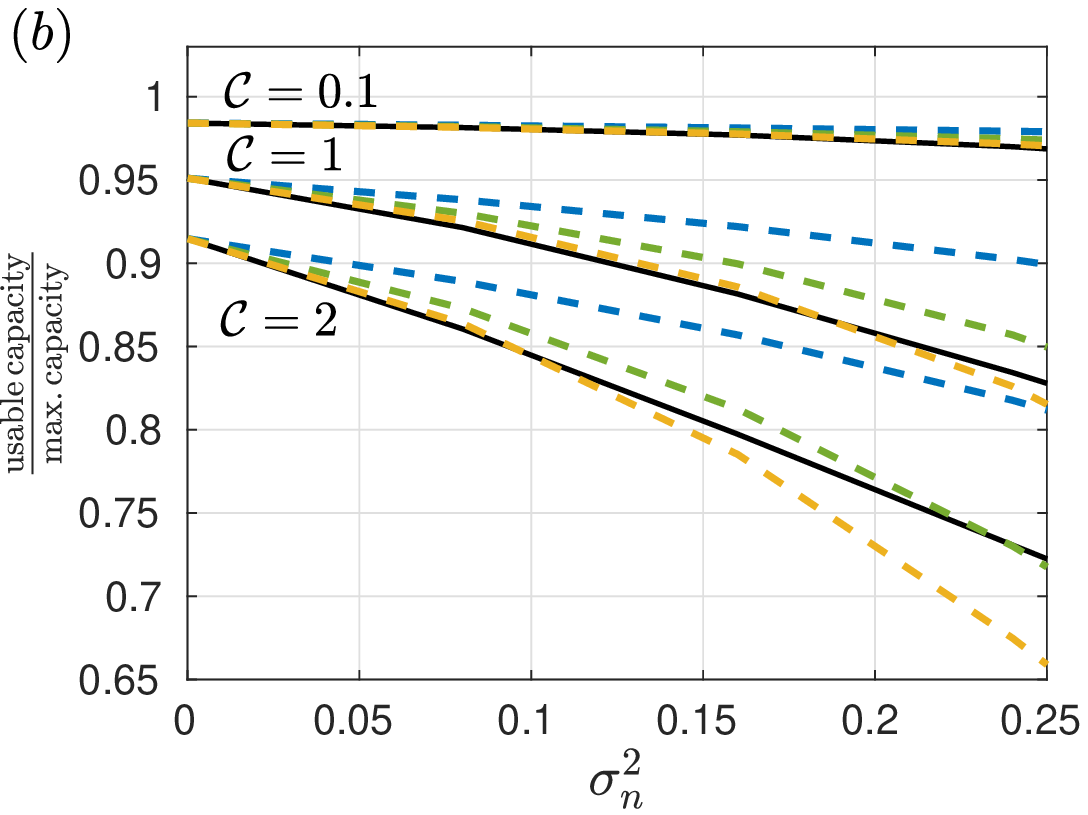}
\par\end{centering}
\caption{$(a)$ Cell potential of the MPM (solid) and several SPMs (dashed)
throughout the discharge with diffusion in the particles, C-rate $\mathcal{C}=1$;
$(b)$ Fraction of maximum theoretical capacity discharged by voltage
cut-off. Mean radii as in caption \ref{fast_diffusion_error_vs_t},
with addition of equivalent capacity radius $\bar{R}_{c}$, given
by (\ref{eq:R_c}).}

\label{diffusion_capacity}
\end{figure}
The potential, and the fraction of the theoretical capacity utilized,
are shown in Fig.~\ref{diffusion_capacity}, for various $\sigma_{n}$.
Increasing $\sigma_{n}$ results in a compression of the discharge
curve horizontally and an earlier arrival at the cut-off voltage.
Hence the greatest capacity
is achieved as $\sigma_{n}\to 0$, i.e. when all the particles are
of the mean radius $\bar{R}_{n}=1$. The capacity effects can be well
reproduced by an SPM with judicious choice of radius. Results for SPMs
at typical mean radii $\bar{R}_{n}$, $\bar{R}_{a}$ and $\bar{R}_{v}$,
along with the newly-derived equivalent-capacity radius $\bar{R}_{c}$
(see \S\ref{subsec:Equivalent-capacity-radius}) are compared
to the full MPM in Fig.~\ref{diffusion_capacity}. The mean radius
itself, $\bar{R}_{n}$, performs the worst, as it is independent of
$\sigma_{n}$ and shows no decrease in capacity (for clarity it is
not included in Fig.~\ref{diffusion_capacity}($b$)). The radii $\bar{R}_{a}$
and $\bar{R}_{v}$ perform better, and were discussed by R\"oder
et al. \cite{Roder2016.}, but the new radius $\bar{R}_{c}$ predicts
capacity the best, and for a wide range of C-rates, despite its derivation being based on a low C-rate. For a given value of $\gamma$, however, the approximation
breaks down as $\sigma_{n}$ increases and progressively larger particles
are included. Nonetheless, use of $\bar{R}_{c}$ is excellent up to
$\sigma_{n}\approx0.5$ if $\mathcal{C}\leq1$ or $\sigma_{n}\approx0.4$
if $\mathcal{C}\leq2$.

One drawback of using the mean radius $\bar{R}_{c}$ is that it underestimates
the surface area compared to the actual PSD. This means, near the
beginning of a discharge, before capacity considerations become important,
$\bar{R}_{c}$ overestimates the reaction resistance and hence the
cell voltage. Here, it is $\bar{R}_{a}$ that best predicts the voltage,
just as for the case of fast diffusion\textemdash see \S\ref{subsec:Results-Fast-diffusion}.

Asymptotic corrections to the  SPMs for narrow distributions
serve to only correct for the shape (i.e. smoothness) of the discharge
curve, exhibiting very similar behaviour (and drawbacks) to those
for fast diffusion, and hence we do not present them here. They do
not correct for errors in capacity due to incorrect choice of radius, highlighting the fact that the choice of effective radius is crucial
when approximation an MPM with an SPM.

\subsubsection{Local concentrations and current densities}
The solution (\ref{eq:Deltaphi_SPM}) for the potential $\Delta\phi^{\mathrm{SPM}}(t)$ 
for an SPM with particle radius $R=R^{\mathrm{SPM}}$ is an estimate for the
potential $\Delta\phi(t)$ of the full MPM. This $\Delta\phi$ is
the same for all particles, hence approximate solutions for the concentrations
in particles of size $R\neq R^{\mathrm{SPM}}$, denoted $c^{(0)}(r,t;R)$,
can be calculated by substituting $\Delta\phi\sim\Delta\phi^{\mathrm{SPM}}$
into (\ref{eq:diffusion_eq})-(\ref{eq:BC_surface}) giving
\begin{align}
\frac{\partial c^{(0)}}{\partial t} & =\frac{\gamma}{r^{2}}\frac{\partial}{\partial r}\left[r^{2}\frac{\partial c^{(0)}}{\partial r}\right], & 0<r & <R,\quad 0<t\label{eq:PSD_1-1-1}\\
-\gamma\frac{\partial c^{(0)}}{\partial r} & =\frac{1}{3}G(c^{(0)},\Delta\phi^{\mathrm{SPM}}), & \text{at }r & =R,\label{eq:PSD_3-1-1}
\end{align}
with $c^{(0)}$ regular at the origin and $c^{(0)}  =c_{\mathrm{init}}$ at
$t  =0$.
There are several advantages to approximating the problem in this
way: (i) each particle size $R$ is decoupled from the others since the integral equation
(\ref{eq:Q_PSD-2}) has been eliminated; in particular  they can be
trivially parallelized; (ii)
the dynamics of particles of any particular size can be investigated
and solved for individually without needing to solve for all particles;
(iii) it allows use of an SPM for battery control and fast online
management, with after-the-fact computation of heterogeneous internal states, and thus nonuniform degradation rates, and at as few (or
many) particle sizes as desired. It has been suggested that variations
in rates of exfoliation and SEI layer growth (common mechanisms of
electrode degradation) from particle to particle are due to local
variations in surface current density \cite{Goers2011}. In nondimensional
terms, this corresponds to $G$, which can be approximated for any
$R$ from the solution of (\ref{eq:PSD_1-1-1})-(\ref{eq:PSD_3-1-1}).

\begin{figure}[htp]
\begin{centering}
\includegraphics[width=0.60\textwidth]{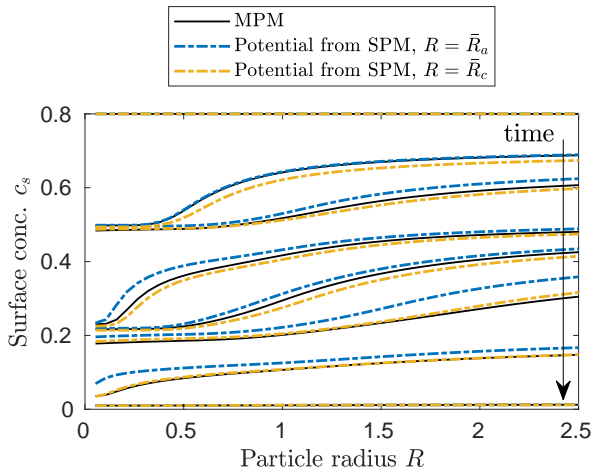}
\par\end{centering}
\caption{Surface concentrations at selected times throughout a discharge, comparing
the full MPM and the solution of (\ref{eq:PSD_1-1-1})-(\ref{eq:PSD_3-1-1})
for two choices of SPM radius: area-weighted mean radius $\bar{R}_{a}$,
and equivalent capacity radius $\bar{R}_{c}$. Here $\mathcal{C}=1$, $\sigma_{n}=0.3$.}
\label{local_concentrations}
\end{figure}

\begin{figure}[htpb]
\begin{centering}
\includegraphics[width=0.68\textwidth]{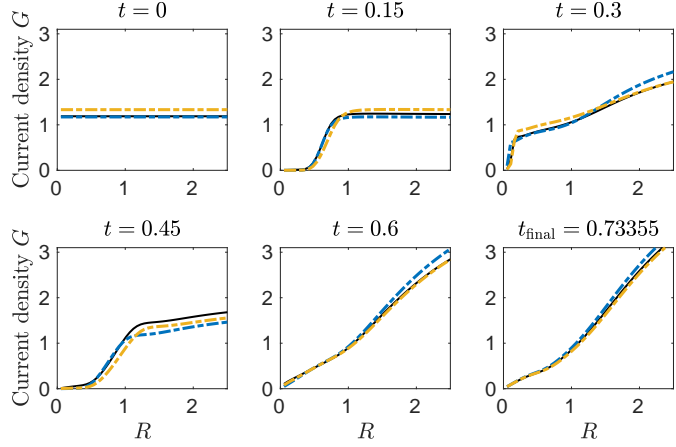}
\par\end{centering}
\caption{Nondimensional local current densities $G$ across particle sizes
at selected times throughout a discharge. See caption and legend of
Fig.~\ref{local_concentrations}.}

\label{local_current_densities}
\end{figure}

The local surface concentrations and current densities as calculated
from (\ref{eq:PSD_1-1-1})-(\ref{eq:PSD_3-1-1}) for $R^{\mathrm{SPM}}=\bar{R}_{a}$
and $R^{\mathrm{SPM}}=\bar{R}_{c}$ are compared to the full MPM solution
in Figs. \ref{local_concentrations} and \ref{local_current_densities}. The approximations using
either radius reproduce the actual concentrations and current densities
surprisingly well for the whole range of particle sizes. Using $\bar{R}_{a}$
is most accurate near the beginning of the discharge and using $\bar{R}_{c}$
is most accurate towards the end, which is expected, since the SPM
potential of each is most accurate at those times. Lastly, it is clear
from Fig.~\ref{local_current_densities} that current densities vary
significantly across particle sizes, suggesting this is an important
heterogeneous effect to capture to predict degradation. In particular,
if, for example,  the cell is cycled over a reduced range of states of charge,
between 100\% and 70\% say, the smaller particles will be used disporportionately,
with larger particles barely used at all in comparison, resulting
in very different degradation rates.

%\pagebreak{}

\section{Bimodal Particle-Size Distributions}
\label{sec:Bimodal-Particle-Size-Distributi}

In this section we will consider PSDs that are bimodal, i.e.~have
two distinct local maxima, such as those in Fig.~\ref{fig:bimodal_1}$(a)$.
These can occur in the manufacturing process from a single source
(see \cite{Rennie2016}) or by the deliberate mixing of two unimodal
distributions from different sources. This latter case of mixing is
especially relevant to the case where two (or more) different electrode
chemistries are mixed, which is becoming common in commercial lithium
cells \cite{Dubarry2011} and has been previously modelled \cite{Albertus2009}.
Here we focus on modelling bimodal distributions of a single chemistry.

A bimodal number density $n(R)$ is modelled as the sum of two unimodal
number densities, $n(R)=n_{1}(R)+n_{2}(R)$, where for each mode $n_{i}(R)$,
$i=1,2$ we  define area densities $a_{i}(R)$, volume densities
$v_{i}(R)$ and the corresponding fraction densities $f_{n,i}(R)$,
$f_{a,i}(R)$, and $f_{v,i}(R)$. We label the modes in order of increasing
mean radii, $\bar{R}_{n,1}<\bar{R}_{n,2}$, and by choice of scaling
we set $\bar{R}_{n,2}=1$. If the total particle number, area and
volume of each mode are $n_{\mathrm{total},i}$, $a_{\mathrm{total},i}$,
$v_{\mathrm{total},i}$, then
\begin{align}
n_{\mathrm{total}} & =n_{\mathrm{total},1}+n_{\mathrm{total},2}, & a_{\mathrm{total}} & =a_{\mathrm{total},1}+a_{\mathrm{total},2}, & 1/3 & =v_{\mathrm{total},1}+v_{\mathrm{total},2},\label{eq:n_a_v_total_bimodal}
\end{align}
where we recall that the total volume of
the PSD has been scaled to be $1/3$. The  distributions of
the mixture are related to those of the individual modes via
\begin{align}
  f_{n}(R) & =\frac{n_{\mathrm{total},1}}{n_{\mathrm{total}}}f_{n,1}(R)+\frac{n_{\mathrm{total},2}}{n_{\mathrm{total}}}f_{n,2}(R),
  \end{align}
and similarly for $f_a$ and $f_v$, where the rational coefficients lie between 0 and 1 and are interpreted
as mixing parameters. If each $f_{n,i}$ is specified, fixing one
coefficient determines the remainder.  We choose to specify
the proportion of the total active material volume contributed by
mode 1, denoted $\delta_{1}=v_{\mathrm{total},1}/(1/3)=3v_{\mathrm{total},1}$.

\subsection{Double particle model (DPM)}

By taking the narrow limit $\sigma_{n,i}\to 0$ as in \S\ref{subsec:Asymptotic-corrections}
for each mode simultaneously,
we can use a single particle radius $R_{i}$ (i.e. any of the means defined in
\S\ref{subsec:Single-particle-models}) to approximate each mode $i$ of the
bimodal PSD, leading to a double
particle model (DPM):
\begin{align}
\frac{\partial c_{i}}{\partial t} & =\gamma\frac{1}{r^{2}}\frac{\partial}{\partial r}\left(r^{2}\frac{\partial c_{i}}{\partial r}\right),&\qquad&\text{for }0<r<R_{i},\label{eq:DPM_diffusion_eq}\\
-\gamma\frac{\partial c_{i}}{\partial r} & =\frac{1}{3}G(c_{i},\Delta\phi)&\qquad&\text{at }r=R_{i},\label{eq:DPM_surface_BC}
\end{align}
for $i\in\{1, 2\}$, with $c_{i}  = c_{\text{init}}$ initially
and $c_i$ regular at $r=0$. These equations are coupled via the algebraic equation
\begin{equation}
-I(t)=a_{\mathrm{total,}1}G(c_{s,1},\Delta\phi)+a_{\mathrm{total},2}G(c_{s,2},\Delta\phi),\label{eq:Q_DPM}
\end{equation}
where the area of each mode, in terms of the volume
share $\delta_{1}$, is
\begin{align}
a_{\mathrm{total},1} & =\frac{\delta_{1}}{R_{1}}, & a_{\mathrm{total},2} & =\frac{1-\delta_{1}}{R_{2}}.\label{eq:a_total_DPM}
\end{align}

\subsubsection{Limit of large mode separation}
 The effects of bimodality
are expected to be the most significant in the limit of large mode
separation, 
$\epsilon=R_{1}/R_{2}\ll1$.
In taking the limit $\epsilon\to0$, we need to decide how the volume share $\delta_1$ scales with $\epsilon$.
 There
are several distinguished limits we could take, for example, $\delta_{1}(\epsilon)=O(\epsilon^{3})$,
$O(\epsilon)$, and $O(1)$, corresponding to fixed particle number
$n_{\mathrm{total},1}$, fixed surface area $a_{\mathrm{total},1}$,
and fixed volume share, respectively. The first two cases give similar
behaviour: the dynamics and cell potential are dominated by mode 2,
with mode 1 providing a small correction of $O(\delta_{1})$. The
last case, where $\delta_{1}=O(1)$ and the volume shares of both
modes are comparable, has a more interesting behaviour and  is
presented below.

Writing $R_{1}=\epsilon R_{2},$ we rescale the domain in particle
1 via the transformation $\tilde{r}=r/\epsilon$, and substitute the
regular expansions
\begin{align}
c_{i} & =\cc_{i}^{(0)}+\epsilon \cc_{i}^{(1)}+O(\epsilon^{2}),\qquad i=1,2,\\
G(c_{i,s},\Delta\phi) & =\GG_{i}^{(0)}+\epsilon\GG_{i}^{(1)}+O(\epsilon^{2}),\qquad i=1,2,\\
\Delta\phi & =\Delta\pphi^{(0)}+\epsilon\Delta\pphi^{(1)}+O(\epsilon^{2}),
\end{align}
in powers of $\epsilon$ into (\ref{eq:DPM_diffusion_eq})-(\ref{eq:Q_DPM}).
The first two orders in (\ref{eq:Q_DPM}) give
\begin{align}
\GG_{1}^{(0)} & =0, \label{eq:G_1_0}\\
\GG_{1}^{(1)} & =-R_{2}I(t)-(1-\delta_{1})\GG_{2}^{(0)},
\end{align}
meaning at leading order particle 1 is in quasi-static equilibrium,
with (\ref{eq:G_1_0}) rearranging to give $\Delta\pphi^{(0)}=U(\cc_{1,s}^{(0)})$, which can be used to eliminate $\Delta\pphi^{(0)}$.
Since particle 1 is small, diffusion is fast there, $\cc^{(0)}_1 = \cc^{(0)}_1(t)$, and the equations simplify as in \S\ref{sec:fastdiffusionmodel}. The analysis is straightforward, and the result is the following coupled system:
\begin{align}
\frac{\mathrm{d}\cc_{1}^{(0)}}{\mathrm{d}t} & =I(t)+\frac{(1-\delta_{1})}{R_{2}}G(\cc_{2,s}^{(0)},U(\cc_{1}^{(0)})),\label{eq:LSL-1}\\
\frac{\partial \cc_{2}^{(0)}}{\partial t} & =\frac{\gamma}{r^{2}}\frac{\partial}{\partial r}\left(r^{2}\frac{\partial \cc_{2}^{(0)}}{\partial r}\right),\qquad\text{for }0<r<R_{2},\label{eq:DPM_diffusion_eq-1}\\
-\gamma\frac{\partial \cc_{2}^{(0)}}{\partial r} & =\frac{1}{3}G(\cc_{2}^{(0)},U(\cc_{1}^{(0)}))\qquad\text{at }r=R_{2}.\label{eq:LSL-2}
\end{align}
with $\cc_{1}^{(0)}  =c_{\mathrm{init}}$, $\cc_{2}^{(0)}  = c_{\text{init}}$
at $t=0$ and $\cc_2$ regular at $r=0$, where as usual $c_{2,s}^{(0)}$ is the surface value $c_2^{(0)}(R_2,t)$. 
 The reduced system (\ref{eq:LSL-1})-(\ref{eq:LSL-2})
is computationally simpler than the DPM (\ref{eq:DPM_diffusion_eq})-(\ref{eq:a_total_DPM}),
and will be compared to the DPM for various size ratios $\epsilon$
in the next section. 

\subsection{Model results}

In this section we present results of the full MPM (\ref{eq:diffusion_eq})-(\ref{eq:Q_PSD-2})
for bimodal PSDs, and the simpler two-particle models: the DPM (\ref{eq:DPM_diffusion_eq})-(\ref{eq:a_total_DPM})
and its large-separation limit (\ref{eq:LSL-1})-(\ref{eq:LSL-2}).
We consider the constant current discharge of a graphite anode, with
the parameters and numerical methods the same as for unimodal distributions
in \S\ref{subsec:Unimodal-Results}. For each mode $i=1,2$
of the PSD, the radius distribution $f_{n,i}(R)$ is a log-normal
with mean $\bar{R}_{i}$ and variance $\sigma_{n,i}^{2}$. 

\subsubsection{Comparison of DPM to full bimodal PSD}

\begin{figure}
\begin{centering}
\includegraphics[width=0.46\textwidth]{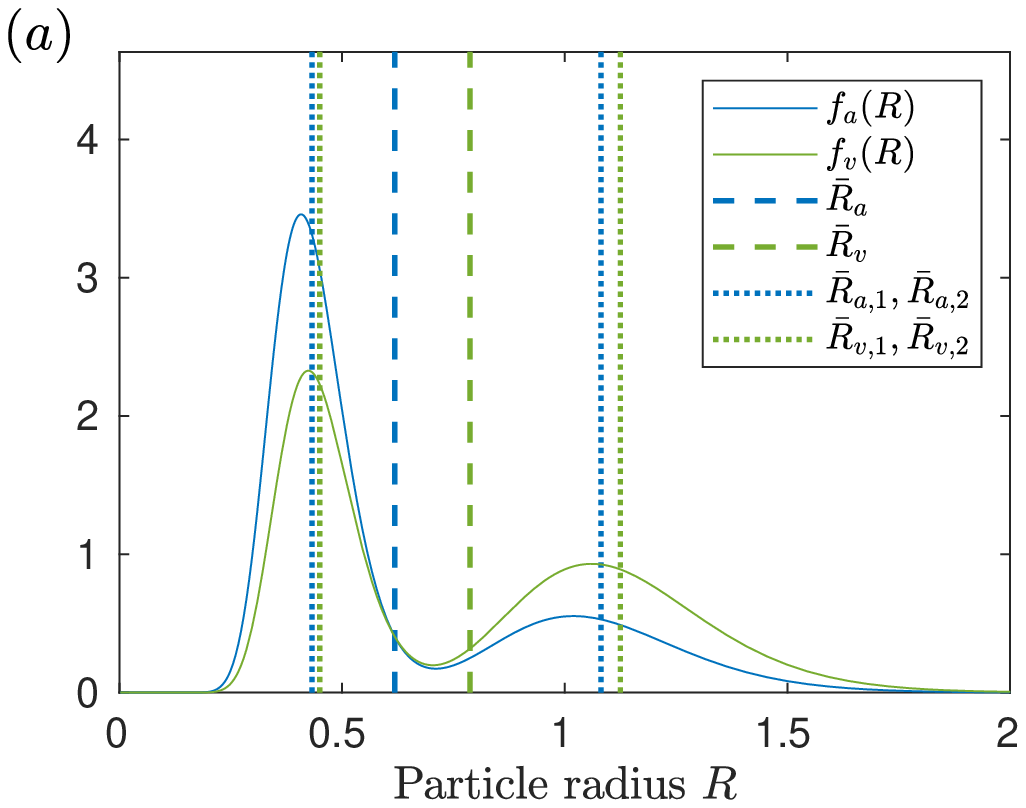}~~~~~\includegraphics[width=0.46\textwidth]{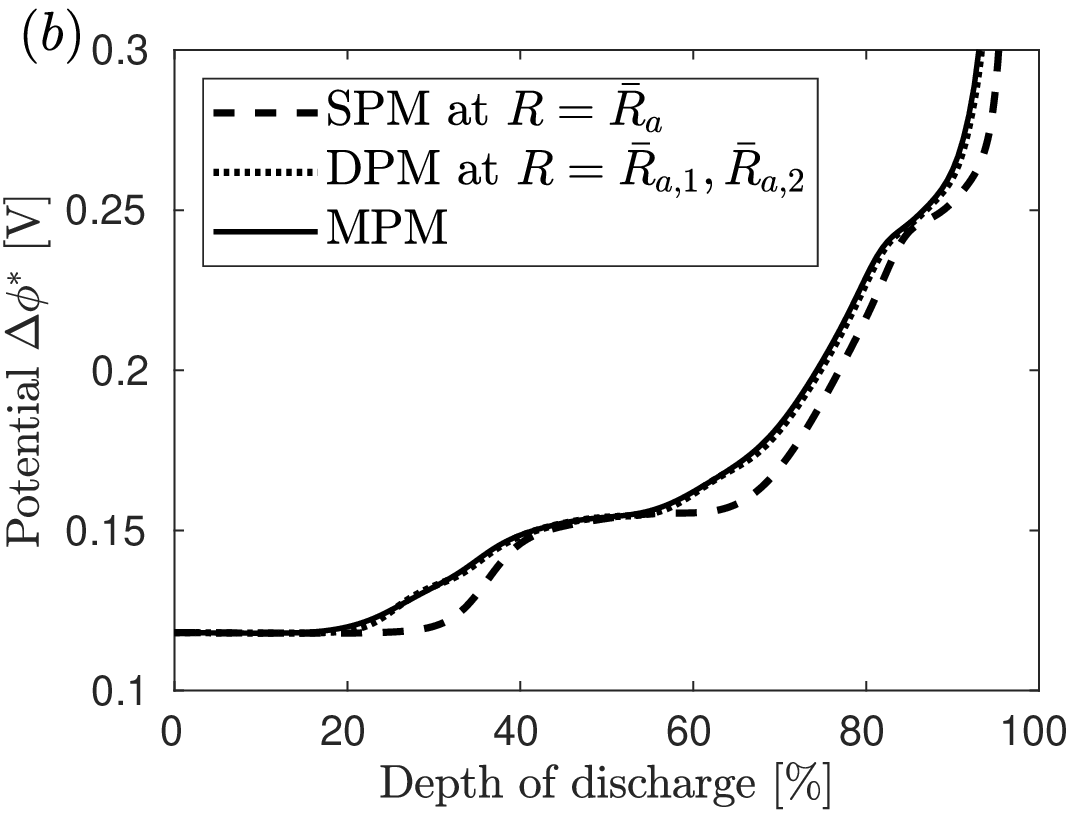}
\par\end{centering}
\caption{\label{fig:bimodal_1}$(a)$ Bimodal PSD, showing area-weighted $f_{a}(R)$
and volume-weighted $f_{v}(R)$ distributions (solid), their means
$\bar{R}_{a}$, $\bar{R}_{v}$ (dashed) and means of each mode (dotted).
$(b)$ Cell potentials for MPM, DPM and SPM based on area-weighted
means. Graphite, parameters $\mathcal{C}=1$, $\bar{R}_{n,1}=0.4$,
$\bar{R}_{n,2}=1$, $\sigma_{a,i}=0.2\bar{R}_{a,i}$, $i=1,2$. }
\end{figure}

\begin{figure}
\begin{centering}
\includegraphics[width=0.5\textwidth]{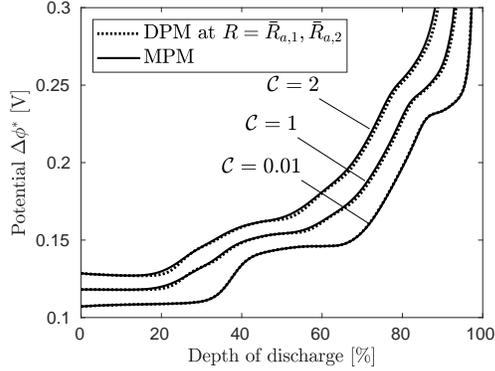}
\par\end{centering}
\caption{\label{fig:bimodal_2}Cell potential for different C-rates $\mathcal{C}$
comparing DPM to MPM for full bimodal PSD, as in Fig.~\ref{fig:bimodal_1}. }
\end{figure}

To assess how well a DPM can reproduce the cell potential of the full
MPM using the bimodal PSD, we choose the radii in the DPM to be the
area-weighted means of each mode, $\bar{R}_{a,i}$, $i=1,2$. We choose the modes to have equal volume share so that $\delta_{1}=0.5$.
The bimodal PSD, the mean radii, and resulting cell potentials are
shown in Fig.~\ref{fig:bimodal_1}. The DPM captures the smoothing of the
steps very well. In contrast, an SPM at the mean
$\bar{R}_{a}$ of the combined bimodal PSD 
cannot capture this effect. Futher, the DPM excellently reproduces the behaviour
with C-rate, shown in Fig.~\ref{fig:bimodal_2}, where the bimodal
nature becomes more apparent as the C-rate increases.

These results justify the use of the DPM in place of the MPM for bimodal
distributions in the remainder of this paper.

\subsubsection{Effect of mode separation}
\label{subsec:Effect-of-mode}
Here we present the impact of mode separation in the DPM.
We fix $\delta_{1}=0.5$ and vary the mode separation through $\epsilon=R_1/R_2$.

\begin{figure}
\begin{centering}
\includegraphics[width=0.43\textwidth]{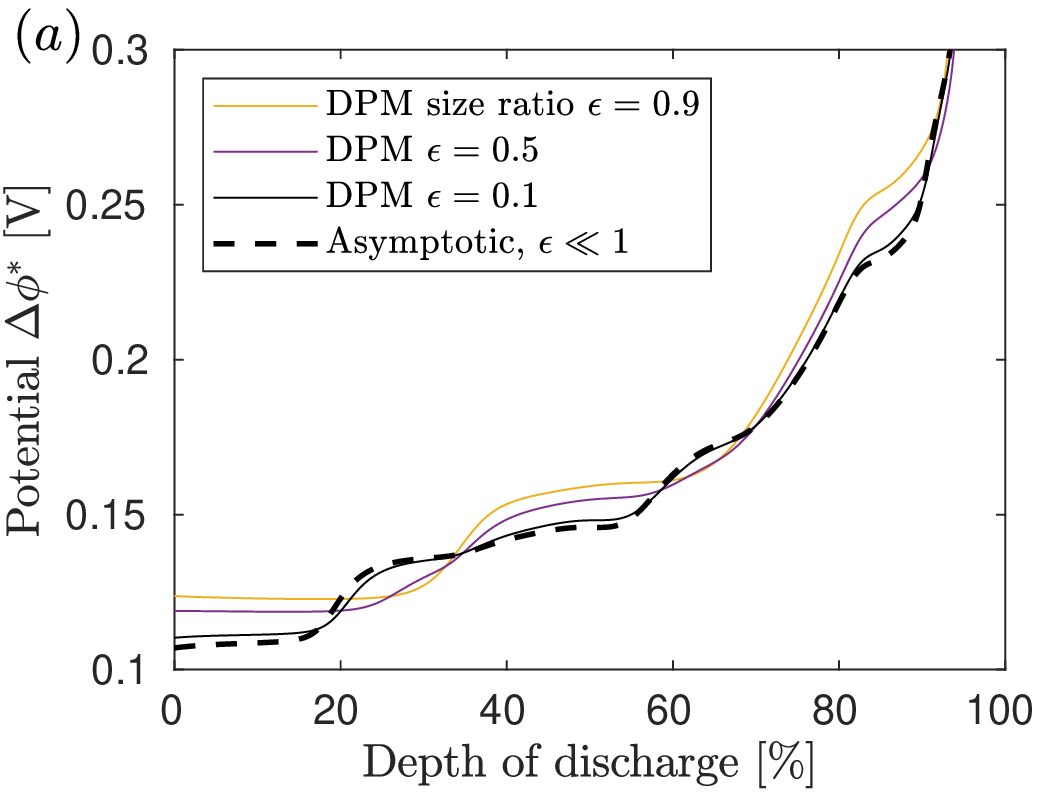}\includegraphics[width=0.43\textwidth]{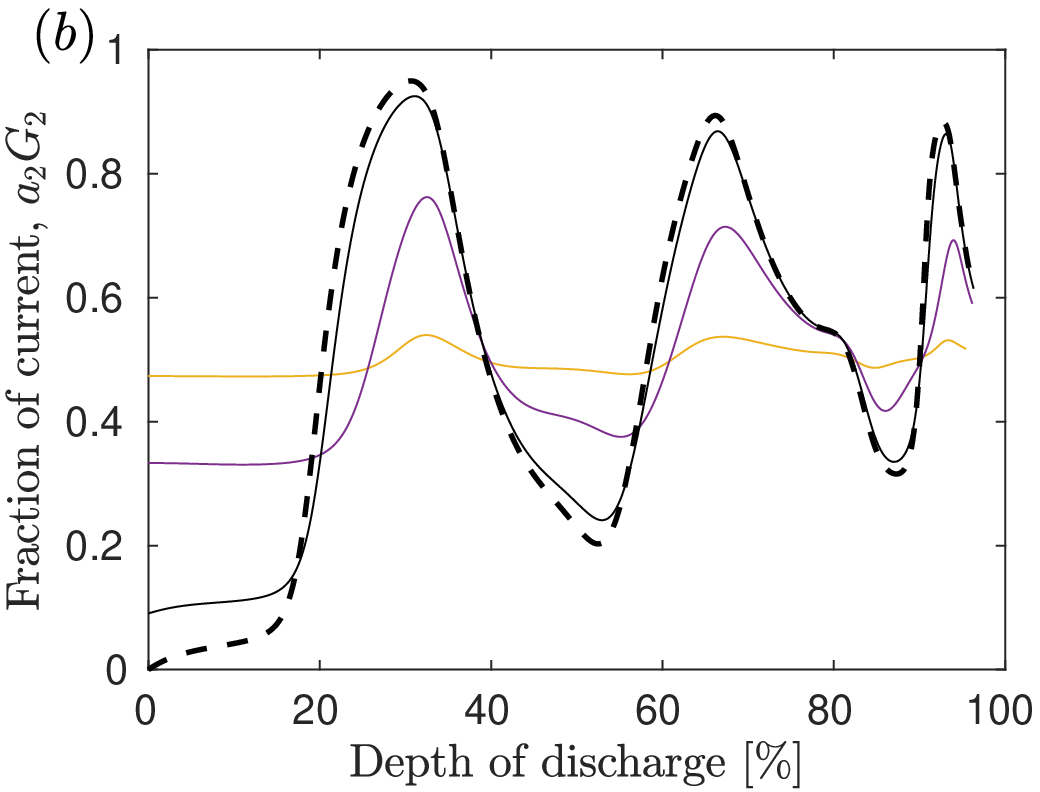}\\
\par\end{centering}
\begin{centering}
\includegraphics[width=0.43\textwidth]{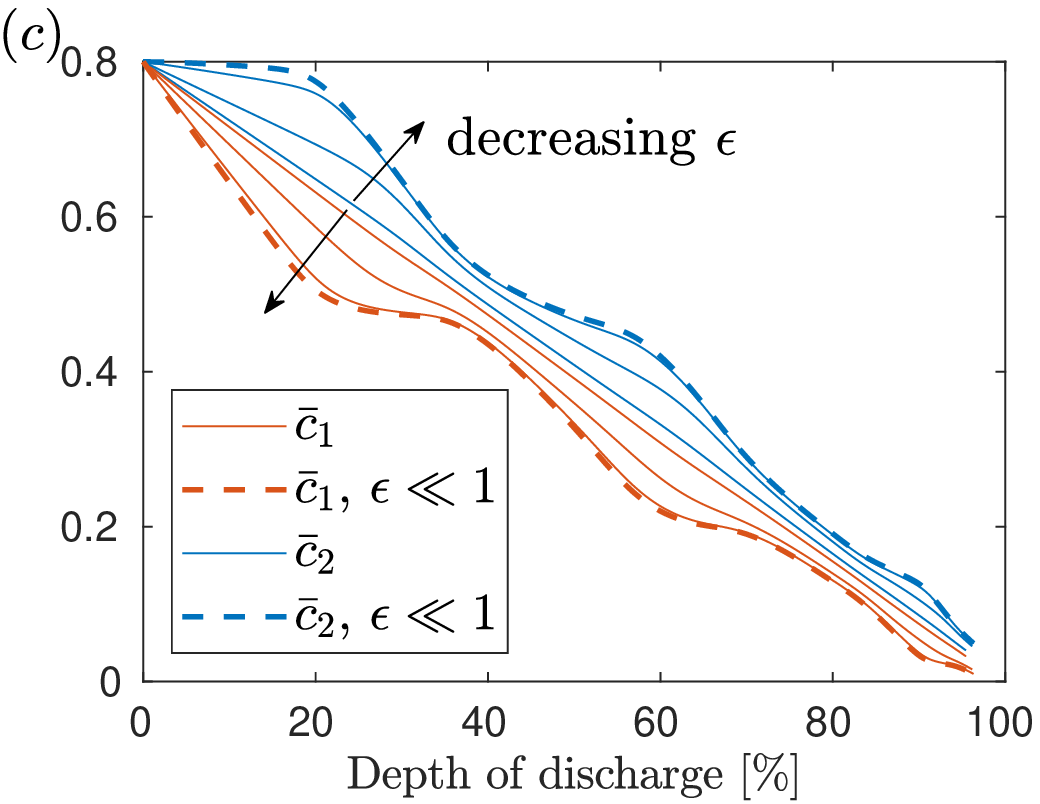}
\par\end{centering}
\caption{\label{fig:bimodal_3}Results of the DPM   with radii $R_{1}=\epsilon$,
$R_{2}=1$, for various size ratios $\epsilon$. Solid lines are the
DPM, dashed are the large separation limit (\ref{eq:LSL-1})-(\ref{eq:LSL-2}). $(a)$ Potential; $(b)$ fraction of current or lithium
flux for (larger) mode 2; $(c)$ particle-averaged concentrations for $\epsilon=0.1$, $0.5$, $0.9$. Graphite, parameters $\delta_{1}=0.5$,
$\mathcal{C}=1$.}
\end{figure}

Results for $\epsilon=0.1,0.5,0.9$, as well as the asymptotic solution
for large separation $\epsilon\ll1$, are shown in Fig.~\ref{fig:bimodal_3}.
When $\epsilon=0.9$, both modes are close to the same size and thus
the results are close to an SPM, and the fraction of current out of
mode 2 (or 1) is approximately a half throughout the discharge\textemdash see
Fig.~\ref{fig:bimodal_3}$(b)$. As $\epsilon$ decreases, the separation
increases, the total surface area of mode 1 increases, but its volume
remains fixed. The result is a staggered depletion of both modes where
the share of the current switches back and forth between the modes
several times due to the ``stepped'' nature of graphite's OCP. The
effect on the potential is nontrivial\textemdash see Fig.~\ref{fig:bimodal_3}$(a)$.
As $\epsilon$ decreases, the plateaus are lowered (decreased reaction
resistance) but the steps are reached earlier as mode 1 depletes faster.

The effects of mode separation are most extreme as $\epsilon\to0$,
where the asymptotic solution for large separation (dashed lines)
is valid. This asymptotic solution agrees well with the solution for
$\epsilon=0.1$, and provides a bound on the effects of bimodality presented
here.

\subsection{Comparison of model to experimental results for LiFePO$_{4}$}
\label{sec4.3}

To further show that the DPM, given by (\ref{eq:DPM_diffusion_eq})-(\ref{eq:a_total_DPM}),
can capture behaviour observed in real cells, we compare the results
of the model to experiments conducted on lithium iron phosphate (LiFePO$_{4}$)
cathodes, a different electrode chemistry to that considered thus far
in this paper. From two different sources of micro-particulate LiFePO$_{4}$
with different mean particle sizes, three cathodes were constructed
consisting of (i) the source of larger particles only (MTI Corp.\textsuperscript{\textregistered});
(ii) the source of smaller particles only (Hydro-Qu\'ebec\textsuperscript{\textregistered});
(iii) a (bimodal) mixture of both particle sizes, in approximately
a 1:1 volume share. Coin half-cells for each type of cathode (denoted
MTI, HQ and MTI+HQ) were constructed and, after minimal cycling, constant
current discharges from 4 V (100\% state of charge) to a cut-off value
of 2.6 V were subsequently performed\textemdash see \S\ref{sec:Experimental-methods} for full details of the experimental procedures.

To parametrise the model we take the electrochemical parameters and
OCP from \cite{Moyles2019}, the total volume fraction $v_{\mathrm{total}}$
and theoretical capacity (and hence the current density $C^{*}$ for
a 1C discharge) from each experimental setup, and then fit the remaining
parameters, i.e., the diffusion coefficient $D^{*}$, radii of each
particle size $R_{1}^{*}<R_{2}^{*}$, and the volume share of the
smaller particles in the mixture, $\delta_{1}$. The cases of only
small or only large particles are taken by setting $\delta_{1}=1$
or $\delta_{1}=0$, respectively. The OCP from \cite{Moyles2019}
was shifted slightly to match the voltage plateau as the C-rate approaches
zero, at 3.409 V. The radii $R_{1}^{*}=120\,\mathrm{nm}$ and $R_{2}^{*}=400\,\mathrm{nm}$
were chosen to fit the centres of the voltage plateaus of the discharge
curves for the small and large particles separately\textemdash comparable
with the manufacturers' estimates of average particle size, $100\,\mathrm{nm}$
and $500\,\mathrm{nm}$. The diffusion coefficient $D^{*}=1\times10^{-17}\mathrm{m}^{2}\text{s}^{-1}$
was chosen to approximate the drop-off of the total discharged capacity
with C-rate. Finally, for the mixture of sizes, the remaining parameter
$\delta_{1}$ was chosen to fit the location of the voltage step,
which is discussed shortly. The complete set of the fitted and experimentally
determined parameters is given in Table \ref{tab:Exp-parameters}.

\begin{figure}
\begin{centering}
\includegraphics[width=0.9\textwidth]{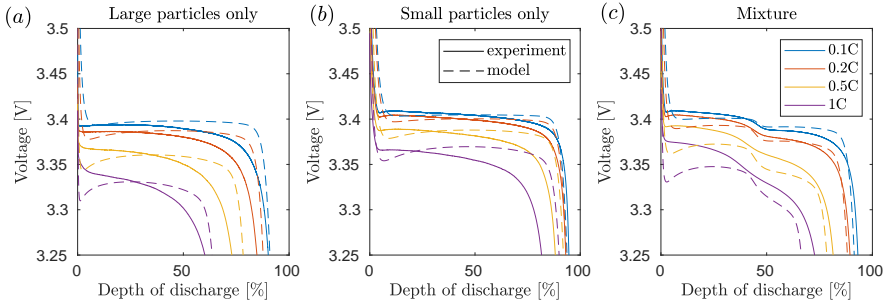}
\par\end{centering}
\caption{\label{fig:Exp-comparison-1}Voltage curves for constant current discharge
of LiFePO$_{4}$ half-cells, comparing experiment to the model (DPM)
 for $(a)$ large particles only (MTI); $(b)$ small particles
only (HQ); ($c$) mixture of small and large particles (MTI+HQ).}
\end{figure}

Discharge voltage curves from the experiments and the model for C-rates
from 0.1 up to 1 are shown in Fig.~\ref{fig:Exp-comparison-1}. When
only one particle size is present, the agreement is only qualitative
as one might expect since, in the model, electrolyte dynamics are
neglected and we assume simple Fickian diffusion of lithium within
the particles, whereas phase change models are more 
appropriate for LiFePO$_{4}$ \cite{Chen2006,Tang2010}. Thus, in
particular, we cannot expect to capture the sloping tail at the end
of discharge for higher C-rates, or the slope of the voltage plateau
itself for lower C-rates (see \cite{Ferguson2014}). Despite this the model is able to reproduce the step between two plateaus which occurs with the
mixture of particles.

\begin{figure}
\begin{centering}
\includegraphics[width=0.9\textwidth]{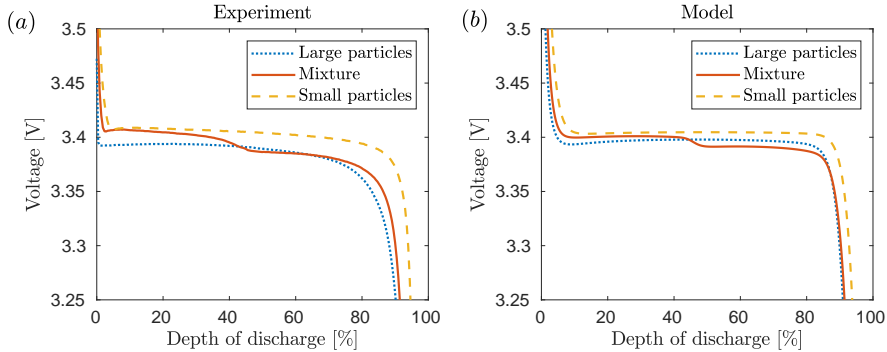}
\par\end{centering}
\caption{\label{fig:Exp-comparison-2}Voltage curves for constant current discharge
of LiFePO$_{4}$ half-cells at 0.1C, with $(a)$ experimental results;
$(b)$ model results (using DPM).}
\end{figure}

This phenomenon is best observed by showing results for all three
electrodes on the same figure, and comparing the differences
between model and experiment, as in Fig.~\ref{fig:Exp-comparison-2}.
In either the model or the experiments, the cases of only smaller
or larger particles  have a single plateau, the former at a higher
voltage due to the lower reaction resistance (higher active surface
area). The mixture shows two plateaus with a step transition midway
through the discharge. In the model this corresponds to intercalation
of smaller particles first until saturation, followed by the larger
particles (just as described for graphite in \S\ref{subsec:Effect-of-mode},
but LiFePO$_{4}$ has only a single plateau). The plateaus of the
mixture are lower than the corresponding ones for a single particle
since the volume share is lower when in the mixture, but the
applied current density is the same. These features are each present
in the experiments, Fig.~\ref{fig:Exp-comparison-2}$(a)$.
A more quantitative
agreement could be achieved by extending the DPM to include electrolyte
dynamics and phase-field physics, but that is outside the scope of
this work. 

Finally, we remark on the differences between this DPM and a similar DPM for spatially (bi-)layered electrodes
presented in Richardson et al. \cite{Richardson2019}. In one scenario considered by \cite{Richardson2019}, the electrode comprises particles of two distinct sizes, separated into two adjacent layers in the through-cell direction. In the asymptotic limit they consider (here corresponding to $\lambda\gg 1$), at leading order they derive a DPM of the form (\ref{eq:DPM_diffusion_eq})-(\ref{eq:a_total_DPM}) with $G_{1}$ and $G_{2}$ elimitated, with the additional restriction that the concentrations on the surface of all particles, regardless of size, are identical for all time. The flux constraint (\ref{eq:DPM_surface_BC}) fixes this concentration, with the potential given by the equilibrium value $\Delta\phi (t)=U(c_{1,s})=U(c_{2,s})$. Since the double-plateau  arises in our model from differing particle surface concentrations, it is not reproduced by the DPM of \cite{Richardson2019}, which is more appropriate for materials whose OCP varies more strongly with state of charge.

\section{Conclusions}
\label{sec:Conclusions}

In this paper we considered heterogeneity of the porous electrodes
in lithium-ion batteries due to the presence of multiple particle
sizes, in particular, particle-size distributions (PSDs).
Using an electrochemical model of a half-cell valid for sufficiently
low C-rates (less than one) where only lithium intercalation
and diffusion within the spherical electrode particles
are  modelled, we presented a detailed investigation of
the effect of the PSD on cell dynamics. We considered both unimodal and
bimodal PSDs,
and various approximations with single and double particle models
(SPM and DPM) were evaluated for different choices of mean particle
radii, of which there are several, with the aim of accurately reducing
the model complexity.

For unimodal distributions, we investigated the effect of the spread
of the PSD for the most common anode material, graphite, where the
two chief effects found were: (1) smoothing of the cell potential
throughout a constant current (dis)charge, a significant effect for
graphite which has a ``stepped'' OCP; (2) reduction in usable capacity
due to the heterogeneous distribution of lithium remaining (or absent)
in different particle sizes at the end of a discharge (or charge).
Results for SPMs for different choices of mean radii were compared
to those for the full PSD, and asymptotic corrections to these models,
consisting of three ``correction particles'', were derived for narrow
PSDs. An SPM can capture effect (2) for a wide range of operating
conditions with a judicious choice of radius, and we systematically
derive a new mean radius for this purpose, (\ref{eq:R_c}), which
depends only on the PSD. An SPM cannot capture the smoothing effect,
(1), but the asymptotic corrections for narrow PSDs can do so up to
$\sigma_{n}^{*}/\bar{R}_{n}^{*}\lesssim0.3$, a physically relevant
range for graphite electrodes. The error in these corrected SPMs is
localised temporally to when the state of the average particle passes
a highly nonlinear section of the OCP. However, we showed that the highly
heterogeneous internal states, e.g. concentrations and current
densities, for all particle sizes can be accurately and efficiently
predicted from an SPM after-the-fact, which may be useful
for predicting nonuniform aging.

Next, we considered bimodal PSDs consisting of a mixture
of two log-normal modes. The dynamics are found to be significantly
different to that of a unimodal PSD but only if the two modes
have a comparable volume share. Then, the cell potential for a full
bimodal PSD is approximated excellently by a double particle model
(DPM) using a single size to represent each mode. For graphite anodes,
the introduction of a second mode has a nontrivial effect,
with the ``stepped'' nature of the OCP resulting in a staggered
(dis)charge where the share of the current switches between modes
several times. An asymptotic limit of the DPM for large mode separation,
eliminating the algebraic equation, gives an upper bound on these
phenomena. Lastly, we presented experimental results for lithium iron
phosphate cathodes with bimodal PSDs formed by mixing two unimodal
PSDs of different means. The results showing a step transition between
two voltage plateaus were reproduced by the DPM despite its mathematical
simplicity and the assumption of linear Fickian solid-state diffusion,
suggesting the phenomenon can be explained entirely by the bimodality
of the PSD.

%\subsection*{Acknowledgements}
%
%TLK, CPP and SJC were supported by the Faraday Institution Multi-Scale
%Modelling (MSM) project, grant number EP/S003053/1.

\appendix

%------------------------------------------------------------
% Body of supplementary material

\section{Parameter values}
\label{sec:Parameter_values_SM}
The tables of parameter values used for the graphite anode and lithium iron phosphate cathode are given here. The dimensional and nondimensional parameters used for the graphite anode are given in Tables \ref{tab:Dimensional-parameters} and \ref{tab:Dimensionless-parameters}, respectively. Those for the lithium iron phosphate cathode used in the DPM to compare with experiment are given in Table \ref{tab:Exp-parameters}.

\begin{table}
\begin{centering}
\begin{tabular}{clc}
\hline 
\multirow{2}{*}{{\footnotesize{}Parameter}} & \multirow{2}{*}{{\footnotesize{}Description}} & {\footnotesize{}Value}\tabularnewline
 &  & {\footnotesize{}\cite{MouraFastDFN,Marquis2019}}\tabularnewline
\hline 
\hline 
$R_{g}^{*}$ & {\footnotesize{}Universal gas constant $[\mathrm{J\,mol}^{-1}\mathrm{K}^{-1}]$} & \multicolumn{1}{c}{8.314472}\tabularnewline
$T^{*}$ & {\footnotesize{}Temperature $[\mathrm{K}]$} & \multicolumn{1}{c}{298.15}\tabularnewline
$F^{*}$ & {\footnotesize{}Faraday's constant $[\mathrm{Cmol}^{-1}]$} & \multicolumn{1}{c}{96487}\tabularnewline
\hline 
$c_{e}^{*}$ & {\footnotesize{}Li concentration in electrolyte $[\mathrm{mol}\,\mathrm{m}^{-3}]$} & $10^{3}$\tabularnewline
$L^{*}$ & {\footnotesize{}Electrode thickness $[\mathrm{m}]$} & $100\times10^{-6}$\tabularnewline
$k^{*}$ & {\footnotesize{}Reaction rate $[\mathrm{A\,m}^{-2}(\mathrm{m}^{3}/\mathrm{mol})^{1.5}]$} & $2\times10^{-5}$\tabularnewline
$c_{\mathrm{max}}^{*}$ & {\footnotesize{}Max. Li concentration in electrode $[\mathrm{mol}\,\mathrm{m}^{-3}]$} & 24983\tabularnewline
$c_{\mathrm{init}}^{*}$ & {\footnotesize{}Initial Li concentration in electrode $[\mathrm{mol}\,\mathrm{m}^{-3}]$} & $0.8c_{\mathrm{max}}^{*}$\tabularnewline
$D^{*}$ & {\footnotesize{}Diffusivity of Li in electrode $[\mathrm{m}^{2}\mathrm{s}^{-1}]$} & $3.9\times10^{-14}$\tabularnewline
$C^{*}$ & {\footnotesize{}Reference current density to discharge in 1hr $[\mathrm{A}\,\mathrm{m}^{-2}]$ } & 24\tabularnewline
$R_{\mathrm{typ}}^{*}$ & {\footnotesize{}Typical particle radius $[\mathrm{m}]$} & $10^{-5}$\tabularnewline
$a_{\mathrm{typ}}^{*}$ & {\footnotesize{}Typical surface area per volume $[\mathrm{m}^{-1}]$} & $1.8\times10^{-5}$\tabularnewline
$v_{\mathrm{total}}$ & {\footnotesize{}Volume fraction of active material} & 0.6\tabularnewline
\hline 
\end{tabular}
\par\end{centering}
\caption{Parameter values for a graphite (LiC$_{6}$) anode \cite{Marquis2019,MouraFastDFN}.}
\label{tab:Dimensional-parameters}
\end{table}

\begin{table}
\begin{centering}
\begin{tabular}{clcc}
\hline 
{\footnotesize{}Nondimensional } & \multirow{2}{*}{{\footnotesize{}Description}} & \multirow{2}{*}{{\footnotesize{}Definition}} & {\footnotesize{}Value}\tabularnewline
{\footnotesize{}parameter} &  &  & {\footnotesize{}\cite{MouraFastDFN,Marquis2019}}\tabularnewline
\hline 
\hline 
$\lambda$ & {\footnotesize{}1 V / thermal voltage} & $F^{*}\Phi^{*}/(R_{g}^{*}T^{*})$ & \multicolumn{1}{c}{38.92}\tabularnewline
$\mathcal{C}$ & {\footnotesize{}C-rate} & $I_{\text{typ}}^{*}/\text{C}^{*}$ & \multicolumn{1}{c}{(variable)}\tabularnewline
\hline 
$k$ & {\footnotesize{}Nondimensional reaction rate} & $\tau_{\text{d}}^{*}/\tau_{\text{reac}}^{*}$ & $7.1103/\mathcal{C}$\tabularnewline
$\gamma$ & {\footnotesize{}Nondimensional diffusivity} & $\tau_{\text{d}}^{*}/\tau_{\mathrm{diff}}^{*}$ & $2.3503/\mathcal{C}$\tabularnewline
$c_{\mathrm{init}}$ & {\footnotesize{}Initial Li concentration} & $c_{\mathrm{init}}^{*}/c_{\mathrm{max}}^{*}$ & 0.8\tabularnewline
\hline 
\end{tabular}
\par\end{centering}
\caption{Nondimensional parameter values for a graphite (LiC$_{6}$) anode.
The dependence of each parameter on the C-rate, $\mathcal{C}$, is
shown explicitly. }
\label{tab:Dimensionless-parameters}
\end{table}

\begin{table}
\begin{centering}
  \resizebox{\textwidth}{!}{
    \begin{tabular}{c>{\raggedright}p{6cm}ccc}
\hline 
\multirow{3}{*}{{\ft Parameter}} & \multirow{3}{*}{{\ft Description}} & \multicolumn{3}{c}{{\ft Experiments}}\tabularnewline
\cline{3-5} 
 &  & {\ft Small particles} & {\ft Large particles} & {\ft Mixture}\tabularnewline
 &  & {\ft (HQ)} & {\ft (MTI)} & {\ft (HQ+MTI)}\tabularnewline
\hline 
$k^{*}$ & {\ft Reaction rate $[\mathrm{A\,m}^{-2}(\mathrm{m}^{3}/\mathrm{mol})^{1.5}]$} & \multicolumn{3}{c}{{\ft $1.35\times10^{-7}$ \cite{Moyles2019}}}\tabularnewline
$c_{\mathrm{max}}^{*}$ & {\ft Max. Li concentration in electrode $[\mathrm{mol}\,\mathrm{m}^{-3}]$} & \multicolumn{3}{c}{{\ft $22806$ \cite{Moyles2019}}}\tabularnewline
$D^{*}$ & {\ft Diffusivity of Li in electrode $[\mathrm{m}^{2}\mathrm{s}^{-1}]$} & \multicolumn{3}{c}{{\ft $1\times10^{-17}$ (fit)}}\tabularnewline
\hline 
$R_{1}^{*}$ & {\ft Particle radius of small particles $[\mathrm{m}]$} & {\ft $120\times10^{-9}$ (fit)} & {\ft -} & {\ft $120\times10^{-9}$}\tabularnewline
$R_{2}^{*}$ & {\ft Particle radius of large particles $[\mathrm{m}]$} & {\ft -} & {\ft $400\times10^{-9}$ (fit)} & {\ft $400\times10^{-9}$}\tabularnewline
$\delta_{1}$ & {\ft Volume share of small particles} & $1$ & $0$ & {\ft $0.4$ (fit)}\tabularnewline
$v_{\mathrm{total}}$ & {\ft Volume fraction of active material} & {\ft $0.3272$} & {\ft $0.3719$} & {\ft $0.3492$}\tabularnewline
$C^{*}$ & {\ft Current density for 1C $[\mathrm{A}\,\mathrm{m}^{-2}]$ } & {\ft $13.60$} & {\ft $14.32$} & {\ft $13.66$}\tabularnewline
\hline 
    \end{tabular}
}
\end{centering}
\caption{Parameter values for LiFePO$_{4}$ cathode used in the DPM (\ref{eq:DPM_diffusion_eq})-(\ref{eq:a_total_DPM}).
Those found by fitting to experiment are indicated.}
\label{tab:Exp-parameters}
\end{table}

\section{Narrow distributions: Asymptotic correction for other mean radii}
 To develop the first correction term 
to an SPM with particle radius $\bar{R}_{i}$, where $i=n,a,v$, we express the
problem in terms of
$f_{i}(R)$.
We can express $a(R)$, which appears in the integral constraint (\ref{eq:Q_PSD-2}),
in terms of any of the radius, area-weighted or volume-weighted distributions
via 
\begin{equation}
a(R)=\begin{cases}
\frac{1}{m_{n,3}}R^{2}f_{n}(R) & \text{(radius)}\\
\frac{1}{\bar{R}_{a}}f_{a}(R) & \text{(area-weighted)}\\
\frac{f_{v}(R)}{R} & \text{(volume-weighted)}
\end{cases}
\end{equation}
The standard deviations $\sigma_{i}$ are a measure of the spread
of each distribution and we can consider the narrow distribution limit
of each by taking $\sigma_{i}\to0$ while fixing $\bar{R}_{i}$. Of
course, the distributions are not independent of each other, and their
moments, $m_{i,j}$, are  related via
\begin{align}
m_{a,j} & =\frac{m_{n,j+2}}{m_{n,2}}, & j & =1,2,...\\
m_{v,j} & =\frac{m_{a,j+1}}{m_{a,1}}=\frac{m_{n,j+3}}{m_{n,3}}, & j & =1,2,...
\end{align}
so that the means and variances satisfy
\begin{align}
\bar{R}_{a} & =m_{a,1}=\frac{m_{n,3}}{m_{n,2}}, & \sigma_{a}^{2} & =m_{a,2}-m_{a,1}^{2}=\frac{m_{n,4}}{m_{n,2}}-\left(\frac{m_{n,3}}{m_{n,2}}\right)^{2}\\
\bar{R}_{v} & =\bar{R}_{a}\left(1+\frac{\sigma_{a}^{2}}{\bar{R}_{a}^{2}}\right), & \sigma_{v}^{2} & =\sigma_{a}^{2}+\frac{\sigma_{a}^{3}}{\bar{R}_{a}}\tilde{\mu}_{3,a}-\frac{\sigma_{a}^{4}}{\bar{R}_{a}^{2}},
\end{align}
where $\tilde{\mu}_{3,a}=\int_{0}^{\infty}(R-\bar{R}_{a})^{3}f_{a}\mathrm{d}R/\sigma_{a}^{3}$
is the coefficient of skewness (standardised third central moment)
representing the asymmetry of $f_{a}$ about its mean. 
It follows from the above relations that if $\sigma_{i}\to0$ for
any $i$, then $\sigma_{n}\sim\sigma_{a}\sim\sigma_{v}\to0$, and
the means are all within $O(\sigma_{i}^{2})$ of each other, e.g.,
$\bar{R}_{v}=\bar{R}_{a}+O(\sigma_{a}^{2})=\bar{R}_{n}+O(\sigma_{n}^{2})$.\footnote{This is true even if we assume each skewness $\gamma_{1,i}$ (and
other higher order moments) is independent of $\sigma_{i}$, or remain
$O(1)$ as $\sigma_{i}\to0$. However, for typical distributions (log-normal,
Weibull), we find that in fact $\gamma_{1,i}=O(\sigma_{i})$ at most.}

In \S\ref{sec3.2} of the paper, the asymptotic limit of narrow distributions
for fixed area-weighted mean radius $\bar{R}_{a}$ was given in detail.
Here we give the corresponding
results for fixed mean radius $\bar{R}_{n}$
and fixed  volume-weighted mean radius $\bar{R}_{v}$.

\subsection{Mean radius}

The integral constraint written in terms of the radius distribution
$f_{n}(R)$, with mean $\bar{R}_{n}$ and variance $\sigma_{n}^{2}$,
is
\begin{equation}
-I(t)=\frac{1}{m_{n,3}}\int_{0}^{\infty}R^{2}f_{n}(R)G(c_{s}(t;R),\Delta\phi(t))\mathrm{d}R.\label{eq:Integral_n}
\end{equation}
The analysis for $\sigma_{n}\ll1$ follows similarly to that of the
area-weighted distribution in \S\ref{sec3.2} of the paper, but with $G$
replaced by $R^{2}G$ in the integral, and with a prefactor of $1/m_{n,3}$.
At $O(\sigma_{n}^{0})$ and $O(\sigma_{n}^{2})$ we find
\begin{align}
-I(t) & =\frac{1}{\bar{R}_{n}}G_{n}^{(0)},\label{eq:G_0_n}\\
 0 & =\bar{R}_{n}^{2}G_{n}^{(2)}+G_{n}^{(0)}+\bar{R}_{n}\left(\frac{\partial G^{(0)}}{\partial R}\right)_{n}+\frac{1}{2}\bar{R}_{n}^{2}\left(\frac{\partial^{2}G^{(0)}}{\partial R^{2}}\right)_{n}-\frac{m_{n,3}^{(2)}}{\bar{R}_{n}^{3}}G_{n}^{(0)},\label{eq:G_2_n}
\end{align}
where we have expanded also $m_{n,3}$ (it depends on $\sigma_{n}$) so that
\begin{align}
\frac{1}{m_{n,3}} & =\frac{1}{m_{n,3}^{(0)}+\sigma_{n}^{2}m_{n,3}^{(2)}+\cdots}=\frac{1}{\bar{R}_{n}^{3}}-\sigma_{n}^{2}\frac{m_{n,3}^{(2)}}{\bar{R}_{n}^{6}}+\cdots.
\end{align}
For
a log-normal $m_{n,3}^{(2)}=3\bar{R}_{n}$. 
At leading order, (\ref{eq:G_0_n}) gives the SPM (\ref{eq:SPM_diffusion_eq})-(\ref{eq:Deltaphi_SPM}) at radius $R=\bar{R}_{n}$, with solution $c_{n}^{(0)}$
and $\Delta\phi^{(0)}$, say.

At $O(\sigma_{n}^{2})$ the correction to the flux  $G_{n}^{(2)}$
depends on both  $\partial G/\partial R$ and $\partial^{2}G/\partial R^{2}$. Then $\partial^{2}G/\partial R^{2}$ is calculated as in \S\ref{sec3.2} (with $\bar{R}_{a}$ replaced by $\bar{R}_{n}$), while $\partial G/\partial R$ can be calculated similarly using
\begin{equation}
\left(\frac{\partial G^{(0)}}{\partial R}\right)_{n}=\begin{cases}
\left(\frac{\partial G^{(0)}}{\partial c}\right)_{n}\left(\frac{\partial c^{(0)}}{\partial R}\right)_{n} &\gamma=\infty,\\
\frac{G(c_{n,+,s}^{(0)},\Delta\phi^{(0)})-G(c_{n,-,s}^{(0)},\Delta\phi^{(0)})}{2\Delta R}+O(\Delta R^{2}) & \text{\ensuremath{\gamma\neq\infty}}.
\end{cases}\label{eq:dGdR_n}
\end{equation}
Then \eqref{eq:PSD_1-1}-\eqref{eq:Deltaphi_SPM-1} still hold (with $\bar{R}_{a}$ and $G_a^{(2)}$  replaced by $\bar{R}_{n}$ and  $G_n^{(2)}$).

\subsection{Volume-weighted mean radius}

The integral constraint written in terms of the volume-weighted distribution
$f_{v}(R)$, with mean $\bar{R}_{v}$ and variance $\sigma_{v}^{2}$,
is
\begin{equation}
-I(t)=\int_{0}^{\infty}f_{v}(R)\frac{G(c_{s}(t;R),\Delta\phi(t))}{R}\mathrm{d}R.\label{eq:Integral_v}
\end{equation}
The analysis for $\sigma_{v}\ll1$ follows similarly to that of the
area-weighted distribution in section \ref{sec3.2}, but
with $G$ replaced by $G/R$ in the integral. At $O(\sigma_{n}^{0})$ and $O(\sigma_{n}^{2})$ we find
\begin{align}
 -I(t) & =\frac{1}{\bar{R}_{v}}G_{v}^{(0)},\label{eq:G_0_v}\\
0 & =\frac{1}{\bar{R}_{v}}G_{v}^{(2)}+\frac{1}{2}\left[\frac{2}{\bar{R}_{v}^{3}}G_{v}^{(0)}-\frac{2}{\bar{R}_{v}^{2}}\left(\frac{\partial G^{(0)}}{\partial R}\right)_{v}+\frac{1}{\bar{R}_{v}}\left(\frac{\partial^{2}G^{(0)}}{\partial R^{2}}\right)_{v}\right].\label{eq:G_2_v}
\end{align}
At leading order, (\ref{eq:G_0_v}) results in the SPM at radius $R=\bar{R}_{v}$,
with solution $c_{v}^{(0)}$ and $\Delta\phi^{(0)}$, say. 

At $O(\sigma_{v}^{2})$, (\ref{eq:G_2_v}) gives the correction $G_{v}^{(2)}$
to the flux. Then $\partial^{2}G/\partial R^{2}$ and $\partial G/\partial R$
are calculated as in \S\ref{sec3.2} (and (\ref{eq:dGdR_n})
above), now at the radius $\bar{R}_{v}$. Given $G_{v}^{(0)}$, \eqref{eq:PSD_1-1}-\eqref{eq:Deltaphi_SPM-1} still hold (with $\bar{R}_{a}$ and $G_a^{(2)}$  replaced by $\bar{R}_{v}$ and  $G_v^{(2)}$).

\section{Expressions for $\partial G/\partial c$ and $\partial^{2}G/\partial c^{2}$}

\label{sec:dGdc_and_d2Gdc2}

In this section we give expressions for the derivatives of the (nondimensional)
surface lithium flux with respect to concentration, for use in \S\ref{sec3.2}. Differentiating the expression \eqref{eq:G_PSD-1} for $G(c,\Delta\phi)$
with respect to $c$,
\begin{align}
\frac{\partial G}{\partial c} & =\frac{g'}{g}G-\frac{\lambda}{2}U'(c)\sqrt{g^{2}+G^{2}},\label{eq:dGdc}\\
\frac{\partial^{2}G}{\partial c^{2}} & =\frac{g''}{g}G+\frac{\lambda^{2}}{4}(U'(c))^{2}G-\frac{\lambda}{2}\sqrt{g^{2}+G^{2}}\left[\frac{2g'(c)}{g(c)}U'(c)+U''(c)\right],\label{eq:d2Gd2c}
\end{align}
where
\begin{align*}
g(c) & =kc^{1/2}(1-c^{1/2}), & g'(c) & =\frac{k^{2}\left(\frac{1}{2}-c\right)}{g}, & g''(c) & =-\frac{\left(k^{2}-(g')^{2}\right)}{g},
\end{align*}
and primes denote $\mathrm{d}/\mathrm{d}c$.
%If evaluated at a leading
%order (i.e., SPM) solution $c_{i}^{(0)}$ and $\Delta\phi^{(0)}$,
%$i=n,a,v$, then $G(c_{i}^{(0)},\Delta\phi^{(0)})=G_{i}^{(0)}=-\bar{R}_{i}I(t)$,
%and (\ref{eq:dGdc})-(\ref{eq:d2Gd2c}) are functions of $c_{i}^{(0)}$
%only.

\section{Experimental methods}

\label{sec:Experimental-methods}

In this section we give the experimental details and methods used
to construct the LiFePO$_{4}$ half-cells referred to in \S\ref{sec4.3}. 

Electrodes were spray deposited from suspensions of LiFePO$_{4}$
with component ratios of 87:3:10 of active material (AM), carboxymethyl
cellulose (CMC), and carbon black (CB), respectively. Slurry mixtures
of the MTI Corp.\textsuperscript{\textregistered} and and HQ (Hydro-Qu\'ebec\textsuperscript{\textregistered}) sources of LiFePO$_{4}$ were prepared at 350 rpm
using 10 mm zirconia balls and a two step mixing process: (1) CB and
2.5 wt. \% CMC aqueous solution were mixed for 15 min and then (2)
LiFePO$_{4}$ and deionised (DI) water were added to the CB-based
slurry and mixed for a further 15 min. In step (2), DI water was added
until the slurry reached a mass concentration of 40 wt. \%. The resulting
slurry was then diluted to a 1 wt. \% solid concentration using DI
water and agitated on a magnetic stirrer for 2 hours before spray
deposition. For the mixed electrode, MT+HQ, a slurry of HQ and MTI
was ball milled, with a HQ:MTI ratio of 47.5:52.5 wt. \%. 

Spray deposition of suspensions was performed with a pneumatic spray
nozzle mounted on an X-Y gantry over a heated vacuum chuck. The temperature
of the substrate was 130 $^{\circ}$C, the flow rate of the suspension
was 4.5 ml/min and the atomisation pressure was 0.4 bar. MTI, HQ and
HQ+MTI mixed suspensions were spray deposited until an electrode thickness
of 80 $\mu$m was deposited. After deposition, the coated aluminium
foils were removed from the heated substrate, calendered and moved
to a vacuum oven at 130 $^{\circ}$C overnight before cell assembly. 

Cells were assembled in a glove box with an Ar atmosphere of $<0.1$
ppm H$_{2}$O and $<0.1$ ppm O$_{2}$. Foil electrodes were assembled
with the following components placed one after another in the centre
of a CR2032 coin cell cup (MTI Corp.): (1) the working electrode,
(2) 75 $\mu$l of 1 molar LiPF$_{6}$ in EC:DMC=1:1 (by volume) electrolyte
(Sigma Aldrich), (3) a glass fibre separator (0.5mm thick, 18mm diameter,
Watson Marlow), (4) 75 $\mu$l of 1 molar LiPF$_{6}$ in EC:DMC=1:1
(by volume) electrolyte, (5) a pre-cut Li chip (0.6 mm thick, 15 mm
diameter, MTI Corp.), (6) a stainless steel spacer (1mm thick, 15.5mm
diameter), (7) a stainless steel wave spring (0.3 mm thick, MTI Corp.)
and (8) a CR2032 coin cell cap (MTI Corp.). The assembled cells were
crimped at 0.08 T (arbitrary units) using an MTI MSK-160E Electric
Crimper and cleaned with ethanol after removal from the glove box.
Cells were allowed to rest for at least 20 hours before initial cycling
was performed.

\bibliographystyle{siamplain}
\bibliography{Refdatabase6SIAP}

\end{document}